%%v12: adding missing case in (11.37)
%%v11: titles of references hidden
%%v10: corrected misprint in 11.17
%%v9 15.2.2015 adding missing case in (11.37)
%%v8 changes on 31.7.12 adding dimensions of N-factors
%%v7 changes in 11.08
%%v6 changes after Proofs 26.4.08
%%v5 changes after referee report 15.12.07
%%small change  20.10.07
%%v4 18.8.2007
%%v3 26.7.2007
\input harvmac
\input amssym.def
\input amssym.tex

\def\tablerule{\noalign{\hrule}}

\parskip=4pt \baselineskip=12pt
\hfuzz=20pt
\parindent 10pt

\def\newsubsec#1{\global\advance\subsecno by1\message{(\secsym\the\subsecno.
#1)} \ifnum\lastpenalty>9000\else\bigbreak\fi
\noindent{\bf\secsym\the\subsecno. #1}\writetoca{\string\quad
{\secsym\the\subsecno.} {#1}}}

\global\newcount\subsubsecno \global\subsubsecno=0
\def\subsubsec#1{\global\advance\subsubsecno
by1\message{(\secsym\the\subsecno.\the\subsubsecno. #1)}
\ifnum\lastpenalty>9000\else\bigbreak\fi
\noindent{\bf\secsym\the\subsecno.\the\subsubsecno.
#1}\writetoca{\string\quad
{\secsym\the\subsecno.\the\subsubsecno.}
{#1}}\par\nobreak\medskip\nobreak}

\def\newsubsubsec#1{\global\advance\subsubsecno
by1\message{(\secsym\the\subsecno.\the\subsubsecno. #1)}
\ifnum\lastpenalty>9000\else\bigbreak\fi
\noindent{\bf\secsym\the\subsecno.\the\subsubsecno.
#1}\writetoca{\string\quad
{\secsym\the\subsecno.\the\subsubsecno.} {#1}}}

\def\nt{\noindent}
\def\nl{\hfill\break}

\def\np{\vfill\eject}

\def\rank{{\rm rank}}
\def\downcirc#1{\mathop{\circ}\limits_{#1}}
\def\riga{-\kern-4pt - \kern-4pt -}
\font\fat=cmsy10 scaled\magstep5

\def\Bbullet{\raise-3pt\hbox{\fat\char"0F}}

\def\black#1{\mathop{\bullet}\limits_{#1}}

\def\mt{\mapsto}
\font\tfont=cmbx12 scaled\magstep1 %large
\font\male=cmr9 

\def\dag{\dagger}

\def\Box{
\vbox{ \halign to5pt{\strut##& \hfil ## \hfil \cr &$\kern -0.5pt
\sqcap$ \cr \noalign{\kern -5pt \hrule} }}~}

\def\down{\raise1.5pt\hbox{$\phantom{a}_2$}\downarrow}

\def\downa{\raise1.5pt\hbox{$\phantom{a}_{2\atop m_2}$}\downarrow}

\def\({\left(}
\def\){\right)}
\def\eps{\epsilon}
 
\def\lra{\longrightarrow}

\def\tL{\tilde{\Lambda}} \def\tl{\tilde{\lambda}}
\def\dia{{$\diamondsuit$}}
\def\lg{\langle} \def\rg{\rangle} 
\def\vf{\varphi}
\def\ha{{\textstyle{1\over2}}}

\def\bac{{C\kern-5.5pt I}}
\def\gc{{\cal G}^{\bac}}

\def\bbz{Z\!\!\!Z}
\def\bbc{{C\kern-6.5pt I}}
\def\bbr{{I\!\!R}}
\def\bbn{I\!\!N}
\def\a{\alpha}
\def\b{\beta}
\def\d{\delta}

\def\vr{\vert}
\def\g{\gamma}
\def\s{\sigma}

\def\l{\lambda}

\def\D{{\Delta}}

\def\ca{{\cal A}} \def\cb{{\cal B}} \def\cc{{\cal C}}
\def\cd{{\cal D}}  \def\cf{{\cal F}}
\def\cg{{\cal G}} \def\ch{{\cal H}} 
 \def\ck{{\cal K}} 
\def\cm{{\cal M}} \def\cn{{\cal N}} 
\def\cp{{\cal P}}  
 \def\ct{{\cal T}}

\def\cih{{\cal C}_\chi}
\def\tcih{{\tilde{\cal C}}_\chi}

\def\th{\theta} \def\Th{\Theta}

\def\tih{{\tilde{\cal T}}^\chi}
\def\tihp{{\tilde{\cal T}}^{\chi'}}

\def\idos{intertwining differential operators}

\def\hc{\ch^\bac}

\def\L{\Lambda}
\def\r{\rho}

\def\nl{\hfill\break}

%%%%%%%%%%%%%%

%%%%%  \nrefs

\nref\BaRo{A.O. Barut and R. R\c aczka, {\it
Theory of Group Representations and Applications}, \hfil\break II
edition, (Polish Sci. Publ., Warsaw, 1980).}

\nref\Ter{J. Terning, {\it Modern Supersymmetry: Dynamics and
Duality}, International Series of Monographs on Physics \# 132,
(Oxford University Press, 2005).}

\nref\Har{Harish-Chandra,
%Discrete series for semisimple Lie groups: II,
Ann. Math. {\bf 116} (1966) 1-111.}

\nref\KnSt{A.W. Knapp and E.M. Stein,
%Intertwining operators for semisimple groups,
Ann. Math. {\bf 93} (1971) 489-578; II : Inv. Math. {\bf 60} (1980)
9-84.}

 \nref\BGG{I.N. Bernstein, I.M. Gel'fand and S.I. Gel'fand,
Funkts. Anal. Prilozh. {\bf 5} (1) (1971) 1-9; English translation:
Funkts. Anal. Appl. {\bf 5} (1971) 1-8.}

\nref\War{G. Warner, {\it Harmonic Analysis on Semi-Simple Lie
Groups I}, (Springer, Berlin, 1972).}

\nref\Lan{R.P. Langlands, {\it On the classification of irreducible
representations of real algebraic groups}, Math. Surveys and
Monographs, Vol.  31 (AMS, 1988), first as IAS Princeton preprint
(1973).}

\nref\Zhea{D.P. Zhelobenko, {\it Harmonic Analysis on Semisimple
Complex Lie Groups}, (Moscow, Nauka, 1974, in Russian).}

\nref\Kosa{B. Kostant,
%Verma modules and the existence of
%quasi-invariant differential operators.
in: Lecture Notes in Math., Vol. 466
(Springer-Verlag, Berlin, 1975) pp. 101-128.}

\nref\Wolfa{J. Wolf, {\it Unitary Representations of Maximal
Parabolic Subgroups of the Classical Groups}, Memoirs Amer. Math.
Soc. 180, (AMS, 1976).}

 \nref\Zheb{D.P. Zhelobenko,
%Discrete symmetry operators for reductive Lie groups
Math. USSR Izv. {\bf 40} (1976) 1055-1083.}

\nref\Dix{J. Dixmier, {\it Enveloping Algebras}, (North Holland, New
York, 1977).}

\nref\Wolfb{J. Wolf, {\it Classification and Fourier inversion for
parabolic subgroups with square integrable nilradical}, Memoirs
Amer. Math. Soc. 225, (AMS, 1979).}

 \nref\KnZu{A.W. Knapp and G.J.
Zuckerman, in: Lecture Notes in Math., Vol. 587 (Springer, Berlin,
1977) pp. 138-159; ~Ann. Math. {\bf 116} (1982) 389-501.}

\nref\DMPPT{V.K. Dobrev, G. Mack, V.B. Petkova, S.G. Petrova and
I.T. Todorov, {\it Harmonic Analysis on the  $n$-Dimensional Lorentz
Group and Its Applications to Conformal Quantum Field Theory},
Lecture Notes in Physics, Vol. 63  (Springer-Verlag,
 Berlin-Heidelberg-New York, 1977).}

\nref\DoPea{V.K. Dobrev and V.B. Petkova, Reports Math. Phys. {\bf
13} (1978) 233-277.}

\nref\SpVo{B. Speh and D.A. Vogan, Jr.,
%Reducibility of generalized principal series representations,
~Acta Math. {\bf 145} (1980) 227-299.}

\nref\Vog{D. Vogan, {\it Representations of Real Reductive Lie
Groups}, Progr. Math.,  Vol. 15 (Boston-Basel-Stuttgart,
Birkh\"auser, 1981).}

\nref\Schl{H. Schlichtkrull,
%A series of unitary irreducible representations induced
%from a symmetric subgroup of a semisimple Lie group,
Inv. Math. {\bf 68}  (1982) 497-516.}

\nref\Speh{B. Speh,
%Unitary representations of Gl(n,R) with
%non-trivial (g,K)-cohomology,
Inv. Math. {\bf 71} (1983) 443-465.}

\nref\Lip{R.L. Lipsman,
%Generic Representations are Induced from
%Square-Integrable Representations,
Trans. AMS  {\bf 285} (1984) 845-854.}

\nref\Dobc{V.K. Dobrev,
%Elementary representations and intertwining operators for\hfil\break $SU(2,2)$: I'',
J. Math. Phys. {\bf 26} (1985) 235-251;
%Multiplet classification of the reducible elementary representations
%of real semi-{\allowbreak}simple Lie groups: the \ $SO_e(p,q)$ example,
V.K. Dobrev, Lett. Math. Phys. {\bf 9} (1985) 205-211.}

\nref\Knaa{A.W. Knapp, {\it Representation Theory of Semisimple
Groups (An Overview Based on Examples)}, (Princeton Univ. Press,
1986).}

\nref\Dob{V.K. Dobrev,
%Canonical construction of intertwining differential operators
%associated with representations of real semisimple Lie groups,
Rep. Math. Phys. {\bf 25} (1988) 159-181; first as ICTP Trieste
preprint IC/86/393 (1986).}

\nref\Jak{H.P. Jakobsen, in: Lect. Notes in Phys., Vol. 261 (1986)
253.}

\nref\BaEa{R.J. Baston and M.G. Eastwood, {\it The Penrose
Transform, Its Interaction with Representation Theory}, (Oxford
Math. Monographs, 1989).}

\nref\GJMS{C.R. Graham et al, %R. Jenne, L. Mason and G. Sparling,
%Conformally invariant powers of the Laplacian, I: Existence,
J. London Math. Soc. (2), 46 (1992) 557-565.}

 \nref\DoMo{V.K. Dobrev  and P. Moylan,
%Induced representations and invariant integral operators for $SU(2,2)$'',
Fort. d. Phys. {\bf 42} (1994) 339-392.}

\nref\Kob{T. Kobayashi,
%Discrete decomposability of the restriction of Aq(?)
%with respect to reductive subgroups and its applications,
Inv. Math. {\bf 117} (1994)  181-205.}

\nref\BOO{T.P. Branson, G. Olafsson and B. Orsted,
%Spectrum Generating Operators, And Intertwining Operators For
%Representations Induced From A Maximal Parabolic Subgroup
J. Funct. Anal. {\bf 135} (1996) 163-205.}

\nref\Eas{M. Eastwood, Suppl. Rend. Circ. Mat.
Palermo, Serie II, Numero 43 (1996) 57-76.}

\nref\DiPe{I. Dimitrov  and I. Penkov,
%Partially integrable highest weight modules,
J. Transf. Groups, {\bf 3} (1998)   241-253.}

\nref\DNW{L. Dolan, C.R. Nappi and E. Witten,
%Conformal Operators for Partially Massless States,
JHEP 0110 (2001) 016, hep-th/0109096.}

\nref\GoMa{X. Gomez  and V. Mazorchuk,
%On an Analogue Of BGG-Reciprocity,
Comm. Algebra, {\bf 29} (2001) 5329-5334.}

\nref\Knab{A.W. Knapp, {\it Lie Groups Beyond an Introduction}, 2nd
ed., Progr. Math., vol. 140 (Boston-Basel-Stuttgart, Birkh\"auser,
2002).}
%1st Edition 1996

\nref\Fra{A. Francis,
%Centralizers of Iwahori-Hecke Algebras II: The General Case,
Alg. Colloquium, {\bf 10}  (2003)   95-100.}

\nref\Kosb{B. Kostant,
%Powers of the Euler product and commutative
%subalgebras of a complex simple Lie algebra,
Inv. Math.  {\bf 158} (2004)   181-226.}

\nref\DFP{I. Dimitrov, V. Futorny  and I. Penkov,
%A Reduction Theorem for Highest Weight Modules over Toroidal Lie Algebras,
Comm. Math. Phys. {\bf 250} (2004) 47-63.}

\nref\Erd{K.  Erdmann, %Type A Hecke algebras and related algebras,
Archiv d. Math. {\bf 82} (2004)   385-390.}

\nref\BaWa{K. Baur and N. Wallach,
%Nice Parabolic Subalgebras of Reductive Lie Algebras,
Represent. Theory, {\bf 9} (2005) 1-29.}

\nref\SaYu{H. Sabourin and R.W.T. Yu,
%On the irreducibility of the commuting variety of a
%symmetric pair associated to a parabolic
%subalgebra with abelian unipotent radical,
%shorter version in:
Journal of Lie Theory {\bf 16} (2006) 57-65; math.RT/0407354.}

\lref\DoPeb{V.K. Dobrev and V.B. Petkova,
%On the group-theoretical
%approach to extended conformal supersymmetry : function space
%realizations and invariant differential operators,
Fortschr. d. Phys. {\bf 35} (1987) 537-572.}

\lref\Dobads{V.K. Dobrev,
%Intertwining operator realization of the AdS/CFT correspondence,
Nucl. Phys. {\bf B553} (1999) 559-582; hep-th/9812194.}

\lref\Dobso{V.K. Dobrev,
%Invariant Differential Operators and Characters of the $AdS_4$ Algebra,
J. Phys. {\bf A39} (2006) 5995-6020; hep-th/0512354.}

%QG

\lref\Dobb{V.K. Dobrev,   Lett. Math. Phys.
{\bf 22} (1991) 251-266; ~V.K. Dobrev,
J. Phys. A: Math. Gen. {\bf 26} (1993)
1317-1334; first as G\"{o}ttingen University preprint, (July 1991);
~V.K. Dobrev and P.J. Moylan,   Phys. Lett. {\bf 315B} (1993)
292-298;  %ICTP Trieste IC/93/059 (March 1993).
~V.K. Dobrev,   J. Phys. A:
Math. Gen. {\bf 27} (1994) 4841-4857 \& 6633-6634, hep-th/9405150.
V.K. Dobrev, Phys. Lett.
{\bf 341B} (1994) 133-138 \& {\bf 346B} (1995) 427.}

\lref\FuMe{V.M. Futorny and D.J. Melville,
%Quantum Deformations of $\a$-Stratified Modules,
Alg. Repres. Theory, {\bf 1} (1998) 135-153.}

\lref\FMJ{F. Fauquant-Millet and A. Joseph,
%Sur les semi-invariants d'une sous-algebre parabolique d'une algebre enveloppante
%quantifiee,
J. Transf. Groups, {\bf 6} (2001) 125-142.}

\lref\JoTo{A. Joseph and D. Todoric,
%On the Quantum KPRV Determinants for Semisimple and Affine Lie Algebras,
Alg. Repr. Theory, {\bf 5} (2002) 57-99.}

\lref\Lya{V.D. Lyakhovsky, Parabolic twists for algebras sl(n),
 math.QA/0510295.}

\lref\BrKl{J. Brundan  and A. Kleshchev,
%Parabolic Presentations of the Yangian,
Comm. Math. Phys. {\bf 254} (2005)  191-220.}

%%\FuMe,\FMJ,\JoTo,\Lya,\BrKl

%integ

\lref\BoFe{J. de Boer and L. Feher,
%Wakimoto Realizations of Current Algebras:
%An Explicit Construction,
Comm. Math. Phys. {\bf 189} (1997) 759-793.}

\lref\GeGi{M. Gerstenhaber  and A. Giaquinto,
%Boundary Solutions of the Classical Yang--Baxter Equation,
Lett. Math. Phys. {\bf 40} (1997) 337-353.}

\lref\DGS{J. Dorfmeister, H. Gradl and J. Szmigielski,
%Systems of PDEs Obtained from Factorization in Loop Groups,
Acta Appl. Math. {\bf 53}  (1998)   1-58.}

\lref\KST{E. Karolinsky, A. Stolin and V. Tarasov,
%From Dynamical to Non-Dynamical Twists,
Lett. Math. Phys. {\bf 71} (2005) 173-178.}

%\BoFe,\GeGi,\DGS,\KST

\lref\Sat{I. Satake, Ann. Math. {\bf 71} (1960) 77-110.}

\lref\Bru{F. Bruhat, Bull. Soc. Math. France, {\bf 84} (1956) 97-205.}

\lref\Dobp{V.K. Dobrev, in preparation.}

\lref\DoPec{V.K. Dobrev and V.B. Petkova,
%All positive energy unitary
%irreducible representations of extended conformal supersymmetry,
Phys. Lett. {\bf 162B} (1985) 127-132;~
%On the group-theoretical approach to extended conformal supersymmetry :
%classification of multiplets,
V.K. Dobrev and V.B. Petkova, Lett. Math. Phys. {\bf 9} (1985)
287-298;~ V.K. Dobrev and V.B. Petkova, Proceedings, eds. A.O. Barut
and H.D. Doebner, Lecture Notes in Physics, Vol. 261
(Springer-Verlag, Berlin, 1986) p. 291-299 and p. 300-308.}

\lref\Dobu{V.K. Dobrev,
%Positive energy unitary irreducible representations
%of D=6 conformal supersymmetry,
J. Phys. {\bf A35} (2002) 7079-7100, hep-th/0201076; ~V.K. Dobrev and R.B. Zhang,
%Positive Energy Unitary
%Irreducible Representations of the Superalgebras $osp(1|2n,R)$,
Phys. Atom. Nuclei, {\bf 68} (2005) 1660-1669,
hep-th/0402039; ~V.K. Dobrev, A.M. Miteva, R.B. Zhang and B.S. Zlatev,
%On the Unitarity of D=9,10,11 Conformal Supersymmetry,
Czech. J. Phys. {\bf 54} (2004) 1249-1256; hep-th/0402056.}

\lref\Dobnt{V.K. Dobrev,
%Characters of the unitarizable highest weight
%modules over the $N=2$ superconformal algebras,
Phys. Lett. {\bf 186B} (1987) 43-51;~
% and ICTP Trieste preprint IC/86/241
%(August 1986).
V.K. Dobrev and A.Ch. Ganchev,
%Modular invariance for the $N=2$ twisted superconformal algebra,
Mod. Phys. Lett. {\bf A3}, 127 (1988).}

\lref\Min{S. Minwalla,
%Restrictions imposed by superconformal invariance on quantum field theories.
Adv. Theor. Math. Phys. {\bf 2}  (1998) 781-846, hep-th/9712074.}

\lref\CCTV{C. Carmeli, G. Cassinelli, A. Toigo and V.S. Varadarajan,
%Unitary representations of super Lie groups and applications to
%the classification and multiplet structure of super particles,
Comm. Math. Phys. {\bf 263}  (2006) 217-258, hep-th/0501061.}

%% START OF TEXT

\centerline{{\tfont Invariant Differential Operators}} \vskip
2truemm \centerline{{\tfont for Non-Compact Lie Groups:}} \vskip
2truemm \centerline{{\tfont Parabolic Subalgebras}}

\vskip 1.5cm

\centerline{{\bf V.K. Dobrev}}
\vskip 0.5cm

\centerline{Institute for Nuclear Research and Nuclear Energy}
\centerline{Bulgarian Academy of Sciences}
\centerline{72 Tsarigradsko Chaussee, 1784 Sofia, Bulgaria}
\centerline{(permanent address)}
%dobrev@inrne.bas.bg

\vskip 0.5cm \centerline{and}

\vskip 0.5cm

 \centerline{The Abdus Salam International Center for
Theoretical Physics} \centerline{P.O. Box 586, Strada Costiera 11}
\centerline{34014 Trieste, Italy}

\vskip 1.5cm

 \centerline{{\bf Abstract}}
\midinsert\narrower{\male In the present paper we start the
systematic explicit construction of invariant differential operators
by giving explicit description of one of the main ingredients - the
cuspidal parabolic subalgebras. We explicate also the maximal
parabolic subalgebras, since these are also important even when they
are not cuspidal. Our approach is easily generalised to the
supersymmetric and quantum group settings and is necessary for
applications to string theory and integrable models.
 }\endinsert

\noindent
%MSC: 17B10, 22E47, 81R05\nl
%PACS: 02.20.Qs, 02.20.Sv, 11.25.Hf

\vskip 1.5cm

\newsec{Introduction}

\nt Invariant differential operators   play very important role in
the description of physical symmetries - starting from the early
occurrences in the  Maxwell, d'Allembert, Dirac, equations, (for
more examples cf., e.g., \BaRo{}),  to the latest applications of
(super-)differential operators in conformal field theory,
supergravity and string theory, (for a recent review, cf. e.g.,
\Ter). Thus, it is important for the applications in physics to
study systematically such operators.

In the present paper we start with the classical situation, with the
representation theory of semisimple Lie groups, where there are lots
of results by both mathematicians and physicists, cf., e.g.
\refs{\Har-\SaYu}. We shall follow a procedure in representation
theory in which such operators appear canonically \Dob{} and which
has been generalized to the supersymmetry setting \DoPeb{} and to
quantum groups \Dobb. We should also mention that
this setting is most appropriate for the classification of unitary
representations of superconformal symmetry in various dimensions,
\DoPec,\Min,\Dobu,\CCTV,
for generalization to the infinite-dimensional setting \Dobnt{}, and is also
an ingredient in the AdS/CFT correspondence, cf. \Dobads. (For a
recent paper with more references cf. \Dobso.)

Although the scheme was developed some time ago there is still
missing explicit description of the building blocks, namely, the
parabolic subgroups and subalgebras from which the representations
are induced.

Just in passing, we shall mention that parabolic subalgebras
found applications in quantum groups, (in particular,
for the quantum deformations of noncompact Lie algebras), cf. e.g.,
\refs{\Dobb,\FuMe,\FMJ,\JoTo,\Lya,\BrKl},
and in integrable systems, cf. e.g.,
\refs{\BoFe,\GeGi,\DGS,\KST}.

In the present paper the focus will be on the role of parabolic
subgroups and subalgebras in representation theory. In the next
section we recall the procedure of \Dob{} and the preliminaries on
parabolic subalgebras. Then, in Sections 3-11 we give the explicit
classification of the cuspidal parabolic subalgebras which are the
relevant ones for our purposes. The cuspidal parabolic subalgebras
are also summarized in table form in an Appendix.

\newsec{Preliminaries}

\newsubsec{General setting}

\nt Let $G$ be a noncompact semisimple Lie group.  Let $K$ denote a
maximal compact subgroup of $G$. Then we have an Iwasawa
decomposition ~$G=KAN$, where ~$A$~ is abelian simply connected, a
vector subgroup of ~$G$, ~$N$~ is a nilpotent simply connected
subgroup of ~$G$~ preserved by the action of ~$A$. Further, let $M$
be the centralizer of $A$ in $K$. Then the subgroup ~$P_0 ~=~ M A
N$~ is a minimal parabolic subgroup of $G$. A parabolic subgroup $P
~=~ M' A' N'$ is any subgroup of $G$ (including $G$ itself) which
contains a minimal parabolic subgroup. The number of non-conjugate
parabolic subgroups is ~$2^r$, where $r=\rank\,A$, cf., e.g., \War.
Note that in general $M'$ is a reductive Lie group with structure:
~$M'=M_d M_s M_a$, where $M_d$ is a finite group, $M_s$ is a
semisimple Lie group, $M_a$ is an abelian Lie group central in $M'$.

The importance of the parabolic subgroups stems from the fact that
the representations induced from them generate all (admissible)
irreducible representations of $G$ \Lan.  (For the role of parabolic
subgroups in the construction of unitary representations we refer to
\Wolfa,\Wolfb.) In fact, it is enough to use only the so-called {\it
cuspidal} parabolic subgroups ~$P=M'A'N'$, singled out by the
condition that ~rank$\, M' =$ rank$\, M'\cap K$ \Zhea,\KnZu, so that
$M'$ has discrete series representations \Har.\foot{The simplest
example of cuspidal parabolic subgroup is $P_0$ when $M'=M$ is
compact. In all other cases $M'$ is non-compact.} However, often
induction from a non-cuspidal parabolic is also convenient.

Let ~$P = M'A'N'$~ be a  parabolic subgroup.
Let ~$\nu$~ be a (non-unitary) character of ~$A'$,
~$\nu\in\ca'^*$, where ~$\ca'$~ is the Lie algebra of $A'$.
If $P$ is cuspidal, let
~$\mu$~ fix a discrete series representation ~$D^\mu$~ of $M'$ on
the Hilbert space ~$V_\mu\,$, or the so-called limit of a discrete
series representation (cf. \Knaa).\foot{In general, $\mu$ is a actually a
triple $(\eps, \sigma, \d)$, where $\eps$ is the signature of the
character of $M_d\,$, ~$\sigma$ gives the unitary character of ~$M_a\,$,
~$\delta$ fixes a discrete or finite-dimensional irrep of $M_s$ on $V_\mu$
(the latter depends only on $\d$).}\nl
Although not strictly necessary, sometimes it is convenient to induce from non-cuspidal $P$
(especially if $P$ is a maximal parabolic). In that case, we use any non-unitary finite-dimensional
irreducible representation ~$D^\mu$~ of $M'$ on the linear space ~$V_\mu\,$.\nl
More than this, except in the case of induction from limits of discrete series,
we can always work with finite-dimensional representations $V_\mu$
by the so-called translation. Namely, when $P$ is non-minimal and cuspidal, then
instead of the inducing discrete series representation of $M'$ we can
consider the finite-dimensional irrep of $M'$ which lies on the same orbit of the Weyl group
(in other words, has the same Casimirs).

We call the induced representation ~$\chi =$ Ind$^G_P(\mu\otimes\nu
\otimes 1)$~ an ~{\it elementary representation} of $G$ \DMPPT.
(These are called {\it generalized principal series representations}
(or limits thereof) in \Knaa.) Their  spaces of functions are:
\eqn\fun{ \cc_\chi ~=~ \{ \cf \in C^\infty(G,V_\mu) ~ \vr ~ \cf
(gman) ~=~ e^{-\nu(H)} \cdot  D^\mu(m^{-1})\, \cf (g) \} } where
~$a= \exp(H)\in A'$, ~$H\in\ca'$, ~$m\in M'$, ~$n\in N'$. ~The
special property of the functions of $\cih$ is called ~{\it right
covariance} \DMPPT,\Dob\ (or {\it equivariance}).\foot{It is well
known that when $V_\mu$ is finite-dimensional ~$\cc_\chi$~ can be
thought of as the space of smooth sections of the homogeneous vector
bundle (called also vector ~$G$-bundle) with base space ~$G/P$~ and
fibre ~$V_\mu\,$, (which is an associated bundle to the principal
~$P$-bundle with total space ~$G$). We shall not need this
description for our purposes.} Because of this covariance the
functions $\cf$ actually do not depend on the elements of the
parabolic subgroup $P=M'A'N'$.

The  elementary representation (ER) ~$\ct^\chi$~ acts in ~$\cc_\chi$
as the left regular representation (LRR) by: \eqn\lrr{
(\ct^\chi(g)\cf) (g') ~=~ \cf (g^{-1}g') ~, \quad g,g'\in G\ .} One
can introduce in ~$\cc_\chi$~ a Fr\'echet space topology or complete
it to a Hilbert space (cf. \War). We shall need also the
infinitesimal version of LRR: \eqn\lac{ ({X_L}\cf) (g) ~\doteq~ {d
\over dt} \cf (\exp(-tX)g)\vr_{t=0} \ , } where, ~$\cf\in \cc_\chi$,
~$g\in G$, ~$X\in \cg$; then we use complex linear extension to
extend the definition to a representation of ~$\gc$.

The ERs differ from the LRR (which is highly reducible) by the
specific representation spaces $\cc_\chi$. In contrast, the ERs are
generically irreducible. The reducible ERs form a  measure zero set
in the space of the representation parameters $\mu$, $\nu$.
(Reducibility here is topological in the sense that there exist
nontrivial (closed) invariant subspace.) The irreducible components
of the ERs (including the irreducible ERs) are called {\it
subrepresentations}.

The other  feature of the ERs which makes them important for our
considerations is a highest (or lowest) weight module structure
associated with them \Dob. For this we shall use the right action of
~$\gc$~ (the complexification of $\cg$) by the standard formula:
\eqn\rac{ (X_R\,\cf) (g) ~\doteq~  {d \over dt} \cf
(g\exp(tX))\vr_{t=0}\ ,  } where ~$\cf\in \cc_\chi$, ~$g\in G$,
~$X\in \cg$; then we use complex linear extension to extend the
definition to a representation of ~$\gc$.  Note that this action
takes ~$\cf$~ out of ~$\cih$~ for some $X$ but that is exactly why
it is used for the construction of the \idos.

We can show this property in all cases when $V_\mu$ is a highest
weight module, e.g., the case of the minimal
parabolic subalgebra and when ~$(M',M'\cap K)$~ is a Hermitian symmetric pair.
In fact, we agreed that,
except when  inducing from limits of discrete series, the space $V_\mu$ will
be finite-dimensional.

Then $V_\mu$ has a highest weight vector $v_0$. Using this we
introduce $\bbc$-valued realization $\tih$ of the space $\cih$ by
the formula: \eqn\sca{ \vf (g) ~\equiv ~ \lg v_0 , \cf(g) \rg ~, }
where ~$\lg , \rg$~ is the $M$-invariant scalar product in $V_\mu$.
~ [If $M'=M_0$ is abelian or discrete then $V_\mu$ is
one--dimensional and ~$\tcih$~ coincides with ~$\cih$; so we set
~$\vf ~=~ \cf$.] ~On these functions the right action of ~$\gc$~ is
defined by:\foot{ In the geometric language we have replaced the
homogeneous vector bundle with base space ~$G/P$~ and fibre
~$V_\mu\,$~ with a line bundle again with base space ~$G/P$ (also
associated to the principal ~$P$-bundle with total space ~$G$). The
functions ~$\vf$~ can be thought of as smooth sections of this line
bundle.} \eqn\raca{ (X_R\, \vf ) (g) ~\equiv ~\lg v_0 , (X_R\, \cf )
(g) \rg ~. }

Part of the main result of our paper \Dob\ is: \hfil\break {\bf
Proposition.} The functions of the ~$\bbc$-valued realization
~$\tih$~ of the ER ~$\cih$~ satisfy : \eqna\low
$$\eqalignno{& X_R\, \vf ~~=~~ \L(X) \cdot \vf ~,
\quad X \in \ch^{\bac} ~, ~~\L \in (\ch^{\bac})^* &\low a\cr &X_R\,
\vf ~~=~~ 0 ~, \quad X \in \gc_+ ~, &\low b\cr }$$ where
%~$\ch' = \ca'\oplus \ch'_{m}$,
%($\ch'_{m}$ being a Cartan subalgebra of $\cm'$, the Lie algebra of $M'$),
 ~$\L= \L(\chi)$ is built canonically from
$\chi\,$,\foot{It contains all the information from $\chi$, except
about the character $\eps$ of the finite group $M_d$. In the case of
G being a complex Lie group we need two weights to encode $\chi$,
cf. Section 3.} ~$\gc_\pm$~ are from the standard triangular
decomposition $\gc = \gc_+ \oplus \ch^{\bac} \oplus
\gc_-\,$.~\foot{Note that we are working here with highest weight
modules instead of the lowest weight modules used in \Dob; also the
weights are shifted by $\r$ with respect to the notation of \Dob.}

Note that conditions \low{} are the defining conditions for the
highest weight vector of a highest weight module (HWM) over $\gc$
with highest weight $\L$. Of course, it is enough to impose \low{b}
for the {\it simple} root vectors ~$X^+_j\,$.

Furthermore, special properties of a class of highest weight
modules, namely, Verma modules, are immediately related with the
construction of invariant differential operators.

To be more specific let us recall that a Verma module is a highest
weight module $V^\L$ with highest weight $\L$, induced from one-dimensional
representations of the Borel subalgebra ~$\cb = \ch^\bac \oplus \gc_+$.
Thus, ~$V^\L \cong
U(\gc_-) v_0$, where $v_0$ is the highest weight vector, $U(\gc_-)$
is the universal enveloping algebra of $\gc_-\,$.~\foot{For more
mathematically precise definition, cf. \Dix.} Verma modules are
universal in the following sense: every irreducible HWM is
isomorphic to a factor-module of the Verma module with the same
highest weight.

Generically, Verma modules are irreducible, however, we shall be
mostly interested in the reducible ones since these are relevant for
the construction of differential equations. We recall the
Bernstein-Gel'fand-Gel'fand \BGG{}  criterion (for semisimple Lie
algebras) according to which the Verma module ~$V^\L$~ is reducible
iff \eqn\vred{ 2 \lg \L + \r , \b \rg ~-~ m \lg \b , \b \rg ~~=~~ 0
~,} holds for some $\b\in\D^+$, $m\in\bbn$, where ~$\D^+$ denotes
the positive roots of the root system $(\gc , \hc)$, $\r$ is half
the sum of the positive roots $\D^+$.

Whenever \vred\ is fulfilled there exists \Dix\ in $V^\L$ a unique
vector $v_s$, called ~{\it singular vector}, which  has the
properties \low{} of a highest weight vector with shifted weight $\L
- m\b$~: \eqna\lows
$$\eqalignno{& X\, v_s ~~=~~ (\L - m\b)(X) \cdot v_s ~,
\quad X \in \ch^{\bac} ~, &\lows a\cr &X\, v_s ~~=~~ 0 ~, \quad X
\in \gc_+ ~, &\lows b\cr}$$

The general structure of a singular vector is \Dob\ : \eqn\siv{ v_s
~~=~~ P_{m\b} (X^-_1,\ldots, X^-_\ell) v_0 ~,} where $P_{m\b}$ is a
homogeneous polynomial in its variables of degrees $mk_i$, where
$k_i \in \bbz_+$ come from the decomposition of $\b$ into simple
roots: $\beta = \sum k_i\alpha_i$, $\alpha_i\in \D_S$, the system of
simple roots, ~$X^-_j$ are the root vectors corresponding to the
negative roots $(-\a_j$), $\a_j$ being the simple roots, $\ell =
\rank_\bac\,\gc =\dim_\bac\,\ch^{\bac}$~ is the (complex) rank of
$\cg^\bac$~.~\foot{A singular vector may also be written in terms
of the full Cartan-Weyl basis of $\gc_-$.}

It is obvious that \siv\ satisfies \lows{a}, while conditions
\lows{b} fix the coefficients of $P_{m\b}$ up to an overall
multiplicative nonzero constant.

Now we are in a position to define the differential intertwining
operators for semisimple Lie groups, corresponding to the singular
vectors.

Let the signature $\chi$ of an ER be such that the corresponding
$\L=\L(\chi)$ satisfies \vred\ for some $\b\in\D^+$ and some
$m\in\bbn$.\foot{If $\b$ is a real root, (i.e., $\b\vr_{\ch^\bac_m}
=0$, where $\ch_m$ is the Cartan subalgebra of $\cm$), then some
conditions are imposed on the character $\eps$ representing the
finite group $M_d$ \SpVo.} Then there exists an intertwining
differential operator \Dob \eqn\intd{ \cd_{m\b} ~:~ \tih ~\lra
~\tihp ~, } where $\chi'$ is such that $\L' = \L'(\chi') = \L -
m\b$.

The most important fact is that \intd\ is explicitly given by \Dob~:
\eqn\ints{ \cd_{m\b} \vf(g) ~~=~~ P_{m\b} ( (X^-_1)_R\,,\ldots,
(X^-_\ell)_R)\, \vf (g) ~,} where ~$P_{m\b}$~ is the same polynomial
as in \siv\ and $(X^-_j)_R$ denotes the action \rac.

One important simplification is that in order to check the
intertwining properties of the operator in \ints\ it is enough to
work with the infinitesimal versions of \fun\ and \lrr, i.e., work
with representations of the Lie algebra. This is important for using
the same approach to superalgebras and quantum groups, and to any other
(infinite-dimensional) (super-)algebra with triangular decomposition.

\newsubsec{Generalities on parabolic subalgebras}

\nt Let $G$ be a real linear connected semisimple Lie
group.\foot{The results are easily extended to real linear reductive
Lie groups with a finite number of components.} Let $\cg$ be the Lie
algebra of $G$, $\th$ be a Cartan involution in $\cg$, and $\cg =
\ck \oplus \cp$ be a Cartan decomposition of $\cg$, so that $\th X
= X, ~X \in \ck$, $\th X = -X , ~X \in \cp$~; $\ck$ is a maximal
compact subalgebra of $\cg$. Let $\ca$ be a maximal subspace of
$\cp$ which is an abelian subalgebra of $\cg$~; $r =\ $dim $\ca$ is
the ~{\it split}~ (or {\it real}) rank of $\cg$, $1 \leq r \leq \ell
= \rank\,\cg$. The subalgebra ~$\ca$~ is called a Cartan subspace of
~$\cp$.

 Let $\D_\ca$ be the root system of the pair $(\cg , \ca)$:
\eqn\rroots{\D_\ca ~\doteq~ \{ \l \in \ca^* ~\vr ~ \l \neq 0 ,\
\cg^\l_\ca \neq 0 \} ~, ~~~\cg^\l_\ca \doteq \{ X \in \cg ~\vr ~
[Y,X] = \l(Y) X ~, ~~\forall Y\in \ca \} ~. } The elements of
$\D_\ca$  are called $\ca\,$-{\it restricted roots}.
 For $\l \in \D_\ca\,$, ~$\cg_\ca^\l$ are called $\ca\,$-{\it\ restricted root
spaces}, ~dim$_R~\cg_\ca^\l \geq 1$.
Next we introduce some ordering (e.g., the lexicographic one) in $\D_\ca\,$.
Accordingly the latter is split into positive   and
negative   restricted roots: ~$\D_\ca = \D_\ca^{+} \cup \D_\ca^{-}$.
Now we can introduce the corresponding nilpotent subalgebras:
\eqn\nilpot{ \cn^\pm ~\doteq~ \mathop{\oplus}\limits_{\l \in \D_\ca^{\pm} }
~~\cg^\l_\ca \ . }

With this data we can introduce the Iwasawa decomposition of ~$\cg$:
\eqn\iwac{ \cg ~=~ \ck \oplus \ca \oplus \cn\ , \quad \cn = \cn^\pm \ . }

Next let $\cm$ be the centralizer of $\ca$ in $\ck$, i.e.,
 $\cm\ \doteq\ \{ X \in \ck ~\vr ~ [X,Y] = 0 , ~\forall Y \in \ca \}$. In
general $\cm$ is a compact reductive Lie algebra, and we can
write $\cm = \cm_s \oplus \cm_a$~, where $\cm_s \doteq [\cm ,
\cm]$~ is the semisimple part of $\cm$, and $\cm_a$ is the
abelian subalgebra central in  $\cm$.

We mention also that  a Cartan subalgebra $\ch_m$ of $\cm$ is given by:
~$\ch_m = \ch_s \oplus \cm_a\,$, where ~$\ch_s$~ is a Cartan subalgebra of $\cm_s\,$.
Then a Cartan subalgebra $\ch$ of $\cg$ is given by:
~$\ch = \ch_m \oplus \ca\,$.\foot{Note that ~$\ch$~ is
a ~$\th$-stable Cartan subalgebra of ~$\cg$~ such that ~$\ch \cap
\cp = \ca$. It is the most noncompact among the non-conjugate
Cartan subalgebras of $\cg$.}

Next we recall the Bruhat decomposition \Bru:
\eqn\bruhat{ \cg ~=~ \cn^+ \oplus \cm \oplus \ca \oplus \cn^-\ ,}
and the subalgebra  ~$\cp_0 \doteq \cm \oplus \ca \oplus \cn^-$~
called a ~{\it minimal parabolic subalgebra}~
of $\cg$. (Note that we may take equivalently
$\cn^+$ instead of $\cn^-\,$.)

Naturally, the $\cg$-subalgebras ~$\ck,\ca,\cn^\pm,\cm,\cm_s,\cm_a,\cp_0$~ are the
Lie algebras of the $G$-subgroups introduced in the previous subsection
~$K,A,N^\pm,M,M_s,M_a,P_0\,$, resp.

We mention an important class of real Lie algebras, the ~{\it split
real forms}. For these we can use the same basis as for the
corresponding complex simple Lie algebra $\cg^\bac$, but over $\bbr$.
Restricting $\bbc\lra\bbr$ one obtains the
Bruhat decomposition of $\cg$ (with $\cm=0$) from  the triangular decomposition of $\cg^\bac
= \cg^+ \oplus \ch^\bac \oplus \cg^-$, and obtains the
   minimal parabolic subalgebras $\cp_0$ from the Borel subalgebra
~$\cb = \ch^\bac \oplus \cg^+$, (or $\cg^-$ instead of  $\cg^+$).
Furthermore, in this case ~$\dim_\bbr\,\ck ~=~ \dim_\bbr\,\cn^\pm$.

   A {\it\ standard parabolic subalgebra}~ is any subalgebra
~$\cp'$~ of $\cg$ containing ~$\cp_0\,$. The number of standard parabolic subalgebras,
including $\cp_0$ and $\cg$, is $2^r$.

\nt {\bf Remark:}~~ In the complex case a standard parabolic subalgebra is any subalgebra
~$\cp'$~ of $\cg^\bac$ containing ~$\cb\,$. The number of standard parabolic subalgebras,
including $\cb$ and $\cg^\bac$, is $2^\ell$,   $\ell = \rank_\bac\ \cg\,$.\dia

Thus, if $r=1$ the only nontrivial parabolic subalgebra is $\cp_0\,$.

Thus, further in this section $r>1$.

Any standard parabolic subalgebra is of the form:
\eqn\parag{\cp' = \cm' \oplus \ca' \oplus \cn'^-\ ,}
so that $\cm' \supseteq \cm$, $\ca' \subseteq \ca$, $\cn'^-
\subseteq \cn^-$~; $\cm'$ is the centralizer of $\ca'$ in $\cg$
(mod~$\ca')$~; $\cn'^-$~   is comprised
from the negative   root spaces of the restricted
root system $\D_{\ca'}$ of $(\cg, \ca')$.
The decomposition \parag{} is called the Langlands decomposition of ~$\cp'\,$.
One also has the analogue of  the Bruhat decomposition  \bruhat:
\eqn\bruhatg{\cg ~=~ \cn'^+ \oplus \ca' \oplus \cm'
\oplus \cn'^- ~, }
where $\cn'^+ = \th \cn'^-$.

The standard parabolic subalgebras may be described explicitly using the restricted simple root system
~$\D^S_\ca\ =\ \D_\ca^+ \cup \D_\ca^-\,$, such that if $\l\in\D_\ca^+$, (resp. $\l\in\D_\ca^-$), one has:
\eqn\decomp{ \l = \sum_{i=1}^r \,n_i\, \l_i\ , ~~~~ \l_i \in\D^S_\ca\ ,
\quad {\rm all}~ n_i\geq 0, ~~~ ({\rm resp.~ all} ~ n_i\leq 0)\ .}

We shall follow Warner \War, where one may find all references to the original
mathematical work on parabolic subalgebras. For a short
formulation one may say that the parabolic subalgebras correspond
to the various subsets of $\D^S_\ca$ - hence their number $2^r$. To
formalize this let us denote: ~$S_r = \{ 1,2,\ldots,r\}$, and  let
~$\Th$~ denote any subset of ~$S_r\,$. Let ~$\D^\pm_\Th\in \D_\ca$~ denote all positive/negative restricted
roots which are linear combinations of the simple restricted roots $\l_i\,, \forall\,i\in \Th$.
Then a standard parabolic subalgebra corresponding to ~$\Th$~ will be denoted by  ~$\cp_\Th$~
and is given explicitly as:
\eqn\stpar{\cp_\Th ~=~ \cp_0\ \oplus\ \cn^+(\Th) \ , \qquad \cn^+(\Th) ~\doteq~
\mathop{\oplus}\limits_{\l\in \Delta^+_\Th}  \ \cg^\l_\ca \ .}
Clearly, ~$\cp_\emptyset ~=~ \cp_0\,$, ~~$\cp_{S_r} ~=~ \cg\,$, ~since
~$\cn^+(\emptyset) ~=~ 0\,$, ~~$\cn^+(S_r) ~=~ \cn^+\,$.
Further, we need to bring \stpar{} in the  form \parag. First, define ~$\cg(\Th)$~ as the algebra generated
by $\cn^+(\Th)$ and $\cn^-(\Th) \doteq \th \cn^+(\Th)\,$.
Next, define $\ca(\Th) \doteq \cg(\Th) \cap \ca$,
and ~$\ca_\Th$~ as the orthogonal complement (relative to the Euclidean structure of $\ca$)
of $\ca(\Th)$ in $\ca$. Then ~$\ca = \ca(\Th) \oplus \ca_\Th\,$.
Note that ~$\dim\, \ca(\Th) = |\Th|$,  ~$\dim\, \ca_\Th = r-|\Th|$.
Next, define:
\eqn\nilp{ \cn^+_\Th ~\doteq~ \mathop{\oplus}\limits_{\l\in \D^+_\ca-\Delta^+_\Th}  \ \cg^\l_\ca\ ,
\quad \cn^-_\Th ~\doteq~ \th\cn^+_\Th \ .}
Then ~$\cn^\pm = \cn^\pm(\Th) \oplus \cn^\pm_\Th\,$.
Next, define ~$\cm_\Th \doteq \cm \oplus \ca(\Th)\oplus \cn^+(\Th) \oplus \cn^-(\Th) $.
Then $\cm_\Th$ is the centralizer of $\ca_\Th$ in $\cg$ (mod $\ca_\Th$).
Finally, we can derive:
\eqn\paragg{\eqalign{ \cp_\Th\ =&\ \cp_0 \oplus \cn^+(\Th) = \cm \oplus \ca \oplus \cn^- \oplus \cn^+(\Th)\ =\cr
=&\ \cm \oplus \ca(\Th) \oplus \ca_\Th \oplus \cn^-(\Th) \oplus \cn^-_\Th \oplus \cn^+(\Th)\ =\cr
=&\ \Big( \cm \oplus \ca(\Th) \oplus \cn^-(\Th) \oplus \cn^+(\Th) \Big) \oplus \ca_\Th \oplus \cn^-_\Th =\cr
=&\ \cm_\Th \oplus \ca_\Th \oplus \cn^-_\Th \ .}}
Thus, we have rewritten explicitly the standard parabolic $\cp_\Th$ in the desired form \parag.
The associated (generalized) Bruhat decomposition \bruhatg{} is given now explicitly as:
\eqn\bruhag{ \eqalign{ \cg\ =&\  \cn^+ \oplus \cp_0 = \cn^+_\Th \oplus \cn^+(\Th) \oplus \cp_0
= \cn^+_\Th \oplus   \cp_\Th \ =\cr =&\  \cn^+_\Th \oplus \cm_\Th \oplus \ca_\Th \oplus \cn^-_\Th \ .}}

Another important class are the {\it maximal\ } parabolic subalgebras which
correspond to $\Th$ of the form:
\eqn\thitb{ \Th^{\rm max}_j ~=~ S_r\backslash \{j\} \ , ~~~1 \leq j\leq r \ .}
~$\dim\, \ca(\Th^{\rm max}_j) = r-1$,  ~$\dim\, \ca_{\Th^{\rm max}_j} = 1$.

\medskip

\nt
{\bf Reminder 1:} ~~~ We recall for further use the fundamental result of Harish-Chandra \Har{}
that $\cg$ has discrete series representations iff ~$\rank\,\cg = \rank\,\ck$.~\dia

\nt
{\bf Reminder 2:} ~~~ We recall for further use the well known fact that
$(\cg,\ck)$ is a Hermitian symmetric pair when the
maximal compact subalgebra  $\ck$ contains a $u(1)$ factor. Then $\cg$ has
highest and lowest weight representations. All these algebras
have discrete series representations.\dia

\newsec{The complex simple Lie algebras considered as real Lie algebras}

\nt
Let ~$\cg_c$~ be a complex simple Lie algebra
of dimension ~$d$~ and (complex) rank ~$\ell$. We need
the triangular decomposition:
\eqn\trig{\cg_c = \cn^+ \oplus \ch \oplus \cn^- \ .}
We have ~$\dim_\bac\, \cg_c = d$, ~$\rank_\bac\,\cg_c = \dim_\bac\, \ch = \ell$,
~$\dim_\bac\, \cn^\pm = (d-\ell)/2$.
Considered as real Lie algebras we have:
~$\dim_\bbr\, \cg_c = 2d$,
~$\rank_\bbr\,\cg_c =\dim_\bbr\, \ch = 2\ell$,
~$\dim_\bbr\, \ck = d$, ~$\rank_\bbr\,\ck =\ell$,
~$\dim_\bbr\, \cn^\pm = d-\ell$.
Note that the maximal compact subalgebra ~$\ck$~ of ~$\cg_c$~
is isomorphic to the compact real form ~$\cg_k$~ of $\cg_c$.

Thus, the complex simple Lie algebras do not have discrete series representations
(and highest/lowest weight representations over $\bbr$).

Let ~$H_j\,$, $j=1,\ldots,\ell$, be a basis of ~$\ch\,$, i.e., ~$\ch
=$~c.l.s.~$\{H_j\,, ~j=1,\ldots,\ell\}$, (where c.l.s. stands for
complex linear span), such that each ad$(H_j)$ has only real
eigenvalues. Let ~$\ca \doteq \ch_{\bbr}\, =$~r.l.s.~$\{H_j\,,
~j=1,\ldots,\ell\}$, where r.l.s. stands for real linear span. Then
the Iwasawa decomposition of ~$\cg_c$~ is: \eqn\iwac{ \cg_c ~=~ \ck
\oplus \ca \oplus \cn\ , \quad \cn = \cn^\pm \ . } The commutant
~$\cm$~ of ~$\ca$~ in ~$\ck$~ is given by: \eqn\mmm{ \cm ~=~ u(1)
\oplus \cdots \oplus u(1)\ , \qquad \ell\ \ {\rm factors} \ .} In
fact, the basis of $\cm$ consists of the vectors $\{i\,H_j\,,
~j=1,\ldots,\ell\}$. The Bruhat decomposition of ~$\cg_c$~ is:
\eqn\bruh{ \cg_c ~=~ \cn^+ \oplus \cm \oplus \ca \oplus  \cn^- \ . }
Comparing \trig{} and \bruh{} we see that \eqn\tribr{ \ch ~=~ \cm
\oplus \ca \ .}

The restricted root system  ~$(\cg_c,\ca)$~ looks the same as the
complex root system ~$(\cg_c,\ch)$, but the restricted roots have multiplicity $2$,
since ~$\dim_\bbr\cn^\pm ~=~ 2 \dim_\bac\cn^\pm$.

Let ~$\Th$~ be a string subset of ~$S_\ell$~ of length ~$s$.
The  ~$\cm_\Th$-factor of the corresponding parabolic subalgebra is:
\eqn\cmcm{ \cm_\Th ~=~ \cg_s \oplus
u(1) \oplus \cdots \oplus u(1)\ , \qquad \ell-s~{\rm factors} \ ,}
where ~$\cg_s$~ is a complex simple Lie algebra of rank $s$
isomorphic to a subalgebra of $\cg_c$.
Thus, the complex simple Lie algebras, considered as real noncompact
Lie algebras, do not have non-minimal cuspidal parabolic subalgebras.

Thus, it is enough to consider elementary representations induced
from the minimal parabolic subgroup ~$P_0 = MAN$, where ~$M \cong
U(1) \times \ldots \times U(1)$, ($\ell$ factors), ~$A \cong SO(1,1)
\times \ldots \times SO(1,1)$, ($\ell$ factors), ~$N \cong \exp
\cn^\pm$.\foot{We should note that the minimal parabolic subgroup
$P_0$ is isomorphic to a Borel subgroup of $G_c$, due to the obvious
isomorphism between the abelian subgroup $MA$ and the Cartan
subgroup $H$ of $G_c$.} Thus, the signature ~$\chi = [\mu,\nu]$,
consists of $\ell$ integer numbers $\mu_i\in\bbz$ giving the unitary
character ~$\mu = (\mu_1\,,\ldots,\mu_\ell)$~ of $M$, and of $\ell$
complex numbers $\nu_i\in\bbc$ giving the character ~$\nu =
(\nu_1\,,\ldots,\nu_\ell)$~ of $A$, ~$\nu_j = \nu (H_j)$. Thus, if $H = \sum_j \s_j H_j$, $\s_j\in\bbr$,
is a generic element of $\ca$, then for the corresponding factor  in \fun{} we have ~$e^{\nu(H)} =
\exp \sum_j \s_j \nu_j\,$. Analogously, if ~$m = \exp i \sum_j \phi_j H_j \in M$, $\phi_j\in\bbr$, then
we have ~$D^\mu(m^{-1}) = \exp i \sum_j \phi_j \mu_j\,$. Thus, the right covariance property \fun{} becomes:
\eqn\funco{\cf (gman) ~=~  \exp \sum_j \(\s_j \nu_j + i \phi_j \mu_j\)\, \cdot\, \cf (g) }

To relate with the general setting of the previous subsection we must introduce two weight
functionals:  ~$\L,\tL$, such that ~$\L(H_j) = \l_j\,$, ~$\tL(H_j) =
\tl_j\,$. Let us use \tribr{} and ~$H = \sum_j (\s_j+i\phi_j) H_j \in\ch\,$.
Thus the elementary representations (in particular, the right covariance
conditions) for a complex semisimple Lie group $G_c$ are given by:
\eqn\fnc{\eqalign{
&\cc_{\L,\tL} ~=~ \{ \cf \in C^\infty(G_c) ~\vr ~\cr &\cf
(gman)  =   \exp\( \L(H) + \tL(\bar{H})\) \cdot  \cf (g) \ = \cr
&=  \exp \sum_j\( (\s_j+i\phi_j) \l_j + (\s_j-i\phi_j) \tl_j \)  \cdot  \cf (g)\  \cr
&\qquad\quad  \nu_j ~=~ \l_j + \tl_j \ ,\quad \mu_j ~=~ \l_j - \tl_j\in \bbz }}
 and the last condition  in \fnc\ stresses that we have
  uniqueness on the compact subgroup ~$M$~ of the Cartan
subgroup ~$H_c =MA$~ of ~$G_c$.

The ERs for which ~$\tL=0$~ are called ~{\it holomorphic}, and those for which
~$\L=0$~ are called ~{\it antiholomorphic}.

Thus, we see that the complex case is richer than the real one.
Indeed, there are ~{\it two\ } Verma modules associated with an ER,
one 'holomorphic' ~$V^\L$~ and one 'antiholomorphic'  ~$V^{\tL}$.
The ER is reducible when either ~$V^\L$~ or ~$V^{\tL}$~ are reducible,
i.e., when \vred\ holds for either $\L$ or $\tL$.

More information  can be found in \Zhea{} from where we mention some
important statements: All irreducible representations of a
complex semisimple Lie group are obtained as subrepresentations of
the elementary representations induced from the minimal parabolic
subgroup. All finite-dimensional irreps   are obtained as subrepresentations
when all ~$\l_j,\tl_j\in\bbz_+\,$.

The maximal parabolic subalgebras have $\cm_\Th$-factors as follows
\eqn\cmcmax{ \cm_\Th ~=~ \cg_i \oplus u(1)  , \qquad i=1,\ldots,\ell \ ,}
where ~$\cg_i$~ is a complex simple Lie algebra of rank $\ell-1$
which  may be obtained from $\cg_c$ by deleting
the $i$-th node of the Dynkin diagram of $\cg_c$.

\newsec{AI :   $SL(n,\bbr)$}

\nt
In this section $G=SL(n,\bbr)$, the group of invertible $n \times n$ matrices with real elements and
determinant 1. Then $\cg = sl(n,\bbr)$ and the Cartan involution is given explicitly by:
~$\th X = -\ ^tX$, where $^tX$ is the transpose of $X\in\cg$. Thus, $\ck \cong so(n)$, and is spanned by matrices
(r.l.s. stands for real linear span):
\eqn\maxc{ \ck = {\rm r.l.s.} \{ X_{ij} \equiv e_{ij} - e_{ji} \ , \quad 1\leq i < j \leq n \} \ ,}
where $e_{ij}$ are the standard matrices with only nonzero entry (=1) on the $i$-th row and $j$-th column,
$(e_{ij})_{k\ell} = \d_{ik}\d_{j\ell}\,$. Note that $\cg$ does not have discrete series representations
if $n>2$. Indeed, the  rank of ~$sl(n,\bbr)$~
is $n-1$, and the rank of its maximal compact subalgebra
$so(n)$ is $[n/2]$ and the latter is smaller than $n-1$ unless
$n=2$.

Further, the complementary space $\cp$ is given by:
\eqn\slp{ \cp = {\rm r.l.s.} \{ Y_{ij} \equiv e_{ij} + e_{ji} \ , ~~ 1\leq i < j \leq n  \ ,
\quad H_j \equiv e_{jj} - e_{j+1,j+1} \ , \quad 1\leq j \leq n-1 \}\ .}
The split rank is ~$r = n-1$, and from \slp{} it is obvious that in this setting one has:
\eqn\aaa{\ca = {\rm r.l.s.} \{ H_j   \ , \quad 1\leq j \leq n-1=r \}\ .}
Since $\cg$ is a maximally split real form of $\cg^\bac = sl(n,\bbc)$, then $\cm=0$, and
the minimal parabolic subalgebra and the Bruhat decomposition, resp.,
are given as  a Borel subalgebra
and triangular decomposition of $\cg^\bac$, but over $\bbr$:
\eqn\brslnr{ \cg ~=~ \cn^+ \oplus  \ca \oplus \cn^-\ , \quad \cp_0 ~=~ \ca \oplus \cn^-\ ,}
where ~$\cn^+,\cn^-$, resp., are upper, lower, triangular, resp.:
\eqn\nslnr{ \cn^+ ~=~ {\rm r.l.s.} \{ e_{ij}\ ,  ~~ 1\leq i < j \leq n\} \ , \quad
\cn^- ~=~ {\rm r.l.s.} \{ e_{ij}\ ,  ~~ 1\leq j < i  \leq n\} \ .}
The simple root vectors are given explicitly by:
\eqn\simple{ X^+_j \doteq e_{j,j+1} \ , ~~~ X^-_j \doteq e_{j+1,j} \ , ~~~ 1\leq j \leq n-1=r \ .}
Note that matters are arranged so that
\eqn\slsub{ [X^+_j,X^-_j] = H_j\ , \quad [H_j,X^\pm_j] = \pm 2X^\pm_j\ ,}
and further we shall denote by ~$sl(2,\bbr)_j$~ the ~$sl(2,\bbr)$~
subalgebra of $\cg$ spanned  by $X^\pm_j\,, H_j\,$.

The parabolic subalgebras may be described by the {\it unordered} partitions of
$n$.\foot{The parabolic subalgebras may also be described by the various flags
of $\bbr^n$, \cf., e.g., \War, but we shall not use this description.}
Explicitly, let ~$\bar \nu \doteq \{\nu_1,\ldots,\nu_s\}$, $s\leq n$, be a partition of $n$:
 ~$\sum_{j=1}^s \nu_j = n$. Then the set $\Th$ corresponding to the  partition  $\bar \nu$
 and denoted by $\Th(\bar \nu)$ consists of the numbers of the entries $\nu_j$
that are bigger than 1: \eqn\thit{ \Th(\bar \nu) ~=~ \{\ j ~|~ \nu_j
> 1\ \} \ .} Note that in the case $s=n$ all $\nu_j$ are equal to 1
- this is the partition ${\bar \nu}_0 = \{1,\ldots,1\}$
corresponding to the empty set: $\Th({\bar \nu}_0) = \emptyset$
(corresponding to the minimal parabolic). Then the factor
$\cm_{\Th(\bar \nu)}$ in \paragg{} and \bruhag{} is: \eqn\pars{
\cm_{\Th(\bar\nu)} ~=~ \mathop{\oplus}\limits_{1\leq j\leq s \atop
\nu_j>1} \ sl(\nu_j,\bbr) ~=~ \mathop{\oplus}\limits_{1\leq j\leq s}
\ sl(\nu_j,\bbr) \ , \quad sl(1,\bbr)\equiv 0 \  }

Certainly, some partitions give isomorphic (though non-conjugate!) $\cm_{\Th(\bar\nu)}$ subalgebras.
The parabolic subalgebras in these cases are called {\it associated}, and this is an equivalence relation.
The parabolic subalgebras up to this equivalence relation correspond to the {\it ordered} partitions of $n$.

The most important for us  cuspidal parabolic subalgebras
correspond to those partitions ~${\bar \nu} =
\{\nu_1,\ldots,\nu_s\}$~ for which $\nu_j \leq 2, ~\forall\,j$.
Indeed, if some $\nu_j>2$ then $\cm_{\Th(\bar\nu)}$ will not have
discrete series representations since it contains the factor
$sl(\nu_j,\bbr)$.

A more explicit description of the cuspidal cases is given as
follows. It is clear that the cuspidal parabolic subalgebras are in
1-to-1 correspondence with the sequences of ~$r$ numbers: \eqn\sequ{
{\bar n} \doteq \{\,n_1,\ldots,n_r\,\} \ ,}  such that $n_j=0,1$,
and if for fixed $j$ we have $n_j=1$, then $n_{j+1}=0$, (clearly
from the latter follows also $n_{j-1}=0$, but we shall use this
notation also in other contexts). In the language above to each
$n_j=1$ there is an entry $\nu_j=2$ in $\bar \nu$ bringing an
$sl(2,\bbr)$ factor to $\cm_\Th\,$, i.e., \eqn\thitz{ \Th(\bar n)
~=~ \{\ j ~|~ n_j = 1 , ~~n_{j+1}=0\ \}\ .} More explicitly, the
cuspidal parabolic subalgebras are given as follows: \eqn\cuspslnr{
\cm_{\Th({\bar n})} ~=~ \mathop{\oplus}\limits_{1\leq t \leq k} \
sl(2,\bbr)_{j_t} \ , \quad n_{j_t}=1, \quad 1\leq j_1 < j_2 < \cdots
< j_k\leq r \ , \quad j_t < j_{t+1}-1 \ .} The corresponding
$\ca_{\Th({\bar n})}$  and $\cn^\pm_{\Th({\bar n})}$ have
dimensions: \eqn\dimanslnr{ \dim\,\ca_{\Th({\bar n})} = n-1-k \ ,
\qquad \dim\,\cn^\pm_{\Th({\bar n})} = \ha n (n-1) -k \ ,} where $k
= \vert \Th({\bar n}) \vert$ was introduced in \cuspslnr.

Note that the minimal parabolic subalgebra is obtained when all ~$n_j=0$,
~${\bar n}_0 = \{0,\ldots,0\}$, then
$\Th({\bar n}_0)=\emptyset$, ~$\cm_{\Th({\bar n}_0)}=0$, ~$k=0$.

\medskip

\nt {\bf Interlude:}~~~ The number of cuspidal parabolic subalgebras of $sl(n,\bbr)$, $n\geq 2$,
including also the case ~$\cp = \cm' = sl(n,\bbr)$~ when $n=2$,
is equal to ~$F(n+1)$, where $F(n)$, $n\in\bbz_+\,$, are the Fibonacci numbers.\nl
{\bf Proof:} ~~
First we recall that the Fibonacci numbers are determined through the relations
~$F(m) = F(m-1) + F(m-2)$, $m\in 2+\bbz_+\,$, together with the boundary values:
$F(0)=0$, $F(1)=1$. We shall count the number of sequences
of ~$r$~ numbers $n_i$, introduced above ($r=n-1$).
Let us denote by $N(r)$ the number of the above-described
sequences. Let us divide these sequences in two groups: the first with $n_1=1$ and the others with
$n_1=0$. Obviously the number of sequences with $n_1=1$ is equal to the $N(r-2)$ since $n_2=0$,
and then we are left with the above-described sequences but of $r-2$ numbers.
Analogously, the number of sequences with $n_1=0$ is equal to the $N(r-1)$ since
 we are left with all above-described sequences   of $r-1$ numbers.
Thus, we have proved that ~$N(r) ~=~ N(r-1)\ +\ N(r-2)$. This is the Fibonacci recursion relation
and we have only to adjust the boundary conditions. We have $N(1)=2$, $N(2) = 3$, i.e.,
~$N(r) = F(r+2)$, or in terms of $n=r+1$: ~$N(n-1) = F(n+1)$.\dia

For further use we recall that there is explicit formula for the
Fibonacci numbers: \eqn\fibo{ F(n) = {x^n - (1-x)^n \over \sqrt5}
~=~ 2^{1-n} \sum_{s=0}^{[(n-1)/2]} {n\choose 2s+1}\ 5^s \quad ,}
where $x$ is the ~{\it golden ratio}~:~ $x^2 =x+1\,$, i.e., ~$x=
(1\pm\sqrt{5})/2\,$.

Finally, we mention that the {\it maximal\ } parabolic subalgebras
corresponding to $\Th$ from \thitb{} have the following factors:
\eqn\cmm{\eqalign{ &\cm_{\Th_j} ~=~  sl(j,\bbr)\oplus sl(n-j,\bbr) \
, ~~~1\leq j\leq n-1\ ,\cr &\dim\,\ca_{\Th_j} ~=~ 1, \quad\dim\,
\cn^\pm_{\Th_j} = j(n-j)}} (Note that the cases ~$j$~ and ~$n-j$~
are isomorphic, or coinciding when ~$n$~ is even and ~$j=\ha n$.)
Only one of the maximal ones is cuspidal, namely, for ~$\cg =
sl(4,\bbr)$, $n=4$~ and $j=2$ we have \eqn\cmtri{\cm_{\Th_2} ~=~
sl(2,\bbr)\oplus sl(2,\bbr) \ . }

\newsec{AII :   $SU^*(2n)$}

\nt The group $G=SU^*(2n)$, $n\geq 2$, consists of all matrices in
$SL(2n,\bbc)$ which commute with a real skew-symmetric matrix times
the complex conjugation operator $C$~: \eqn\SUs{SU^*(2n) \doteq \{\
g\in SL(2n,\bbc) ~|~ J_n C g= g J_n C \ , ~~~J_n \equiv \pmatrix{0 &
1_n  \cr -1_n & 0 } \    \} \ .} The Lie algebra $\cg=su^*(2n)$ is
given by: \eqn\sus{\eqalign{ su^*(2n) &\doteq \{\ X\in sl(2n,\bbc)
~|~ J_n C X= X J_n C  \ \} \ = \cr &= \ \{\ X= \pmatrix{a & b \cr -
{\bar b} & {\bar a}} ~|~\ a,b\in gl(n,\bbc)\ , ~~~  {\rm tr}\,
(a+{\bar a})=0 \ \}\ .}} $\dim_R\,\cg = 4n^2-1$.

We consider $n\geq 2$ since $su^*(2) \cong su(2)$,
and we note that the case $n=2$ (of split rank 1) will appear also below:
 ~$su^*(4) \cong so(5,1)$, cf. the corresponding Section.

The Cartan involution is given   by: ~$\th X = -X^\dag$. Thus, $\ck
\cong sp(n)$: \eqn\susc{\ck = \{\ X= \pmatrix{a & b \cr -b^\dag & -\
^ta} ~|~\ a,b\in gl(n,\bbc) \ , ~~~a^\dag = -a \ , ~~~ ^tb= b  \ \}\
. } Note that $su^*(2n)$ does not have discrete series
representations (rank$\,\ck = n <$ rank$\,su^*(2n)=2n-1$). The
complimentary space $\cp$ is given by: \eqn\susc{\cp = \{\ X=
\pmatrix{a & b \cr b^\dag & ^ta} ~|~\ a,b\in gl(n,\bbc) \ ,
~~~a^\dag = a \ , ~~~ ^tb= -b, ~~~{\rm tr}\,  a=0  \ \}\ . } The
split rank is ~$n-1$~ and the abelian subalgebra $\ca$ is given
explicitly by: \eqn\susa{ \ca = \{\ X = \pmatrix{a & 0 \cr 0 & a}
~|~ a =  {\rm diag}\, (a_1,\ldots,a_n), ~~~a_j\in\bbr \ ,~~~~{\rm
tr}\, a  =0 \ \}\ . } The subalgebras ~$\cn^\pm$~ which form the
root spaces of the root system $(\cg,\ca)$ are of real dimension
~$2n(n-1)$. The subalgebra $\cm$ is given by: \eqn\susm{\eqalign{
\cm  ~=&~ \{\ X = \pmatrix{a & b \cr -{\bar b} & -a} ~|~ a = i\,
{\rm diag}\, (\phi_1,\ldots,\phi_n), ~~\phi_j\in\bbr \ , \cr
 &\qquad b = {\rm diag}\, (b_1,\ldots,b_n), ~~b_j\in\bbc\ \} \ \cong \cr
\cong& ~ su(2) \oplus\cdots \oplus su(2) \ , ~~n~{\rm factors} \ .}}

\nt
{\bf Claim:} ~~~All non-minimal parabolic subalgebras of $su^*(2n)$ are not cuspidal.\nl
{\bf Proof:} ~~~Necessarily $n>2$. Let $\Th$ enumerate a connected string of restricted simple roots:
~$\Th = S_{ij} = \{\ i,\ldots,j\ \}$, where $1\leq i\leq j <n$.
Then the corresponding subalgebra $\cm_\Th$ is given by:
\eqn\cmsumm{\cm_{ij} = su^*(2(s+1)) \oplus su(2) \oplus\cdots \oplus su(2) \ ,
~~n-s-1~{\rm factors} \ , ~~~s\equiv j-i+1 \ .}
In general $\Th$ consists of such strings, each string of length $s$  produces
a factor $su^*(2(s+1))$, the rest of $\cm_\Th$ consists of $su(2)$ factors.\dia

The maximal parabolic subalgebras, cf. \thitb{}, $1\leq j \leq n-1$,
contain   $\cm_\Th$ subalgebras of the form:
\eqn\cmsmax{ \cm^{\rm max}_j ~=~ su^*(2j) \oplus su^*(2(n-j))  \ .  }
(For $j=1$ or $j=n-1$  \cmsmax{} coincides with \cmsumm{} for $s=n-2$
(and using $su^*(2) \cong su(2)$).

\newsec{AIII,AIV :  $SU(p,r)$}

\nt
In this section $G=SU(p,r)$, $p\geq r$, which standardly is defined as
follows: \eqn\SUpr{SU(p,r) \doteq \{\ g\in GL(p+r,\bbc) ~|~ g^\dag
\b_0 g = \b_0 \ , ~~~\b_0 \equiv \pmatrix{1_p & 0 \cr 0 & -1_r
} \ , ~~~\det\,g =1\  \} \ ,} where $g^\dag$ is the Hermitian
conjugate of $g$.   We shall use also another
realization of $G$ differing from \SUpr{} by unitary transformation:
\eqn\SUppr{\b_0 \mt \b_2 \equiv \pmatrix{1_{p-r}& 0 & 0\cr 0 & 0 & 1_r \cr 0& 1_r & 0 } =
U\b_0 U^{-1} \ , \quad U \equiv {1\over\sqrt{2}} \pmatrix{1_{p-r} & 0 & 0\cr 0 & 1_r & 1_r
\cr 0& -1_r & 1_r} \ .} The Lie algebra $\cg=su(p,r)$ is given by
($\b=\b_0,\b_2$): \eqn\supr{ su(p,r) \doteq \{\ X\in gl(p+r,\bbc) ~|~
X^\dag \b + \b\,X = 0 \ , ~~~ {\rm tr}\, X =0 \ \}\ . } The Cartan
involution is given explicitly by: ~$\th X = \b X\b$. Thus, $\ck
\cong u(1) \oplus su(p)\ \oplus su(r)$, and more explicitly is
given as ($\b = \b_0$): \eqn\maxcs{ \ck =  \{\ X = \pmatrix{u_1 &
0 \cr 0 & u_2} ~|~ u^\dag_j = -u, ~~~j=1,2; ~~~~{\rm tr}\,
u_1\ +\ {\rm tr}\, u_2  =0 \ \}\ . }

Note that $su(p,r)$  has discrete series representations since
~$\rank\, \ck = 1 +\rank\, su(p) + \rank\, su(r) = p+r-1  = \rank\, su(p,r)$,
and highest/lowest weight representations.

The split rank is equal to ~$r$~ and the abelian subalgebra $\ca$ may
be given explicitly by ($\b = \b_2$): \eqn\aaa{\ca = {\rm r.l.s.} \{\
H^u_j   \equiv e_{p-r+j,p-r+j} - e_{p+j,p+j} \ , \quad 1\leq j \leq r\ \}\ .
}

At this moment we need to consider the cases $p=r$ and $p>r$ separately,
since the minimal parabolic subalgebras are different.

\newsubsec{The case  $SU(n,n),$ $n>1$}

\nt
In this subsection $G=SU(n,n)$.  We consider $n>1$ since $SU(1,1) \cong
SL(2,\bbr)$, which was already treated.

The subalgebra $\cm \cong u(1) \oplus \cdots\oplus u(1)$, (${n-1}$ factors),
and is explicitly given as ($\b = \b_2$):
\eqn\maxcm{ \cm =  \{\ X = \pmatrix{u & 0 \cr 0 & u} ~|~ u = i\, {\rm diag}\,
(\phi_1,\ldots,\phi_n), ~~~\phi_j\in\bbr \ ; ~~~~{\rm tr}\, u  =0 \ \}\ . }
The subalgebras ~$\cn^\pm$~ which form the root spaces of the root system $(\cg,\ca)$
are of real dimension ~$n(2n-1)$. The simple root system $(\cg,\ca)$ looks as that of the
symplectic algebra $C_n$, however, the root spaces of the short roots have multiplicity 2.

Further, we choose the long root of the $C_n$ simple root system to be $\a_n\,$.

{\bf Claim:}~~ The nontrivial cuspidal parabolic subalgebras are given by $\Th$ of the form:
\eqn\thita{ \Th_j ~=~ \{\ j+1, \ldots, n\ \}\ , ~~~1 \leq j < n \ .}
{\bf Proof:}~~ First note that we exclude $j=0$ since $\Th_0 = S_n\,$. Consider now any $\Th$
which contains a subset $S_{ij} = \{\ i,\ldots,j\ \}$, where $1\leq i\leq j <n$.
  Then the simple roots corresponding to $S_{ij}$ form a string subset $\pi_{ij}$ of the
simple root system of $A_{n-1}\,$, but each root has multiplicity 2.  Because of this multiplicity
this  string of simple roots   will produce a subalgebra $sl(j-i+2,\bbc)$ of
 $\cm_\Th$. Since the simple Lie algebras $sl(n,\bbc)$ do not have discrete series
 representations, then $\cp_\Th$ is not cuspidal. Now it remains to note that $\cp_{\Th_j}$
is cuspidal for all $j$ since
\eqn\cmmu{ \cm_{\Th_j} ~\cong~ su(n-j,n-j) \oplus
u(1) \oplus \cdots\oplus  u(1), ~~~j~ {\rm factors}  }
cf. the Remark above.\dia

The maximal parabolic subalgebras correspond to the sets ~$\Th^{\rm max}_j\,$,
$j=1,\ldots,n$, cf. \thitb. The corresponding $\cm_\Th$
subalgebras are of the form:
\eqn\cmmax{ \cm^{\rm max}_j ~=~ sl(j,\bbc) \oplus su(n-j,n-j)\oplus  u(1)
\oplus \cdots\oplus  u(1), ~~~j~ {\rm factors} \ ,  }
where we use the convention: $sl(1,\bbc) =0$.
Note that $\Th^{\rm max}_1 = \Th_1$ and that
the only cuspidal maximal parabolic subalgebra is ~$\cp_{\Th_1}\,$.\nl
The latter is also the only Heisenberg parabolic subalgebra. %25.9.2021

\newsubsec{The case  $SU(p,r)$, $p>r\geq 1$}

\nt
In this subsection $G=SU(p,r)$. We include also the case $r=1$ although
we noted that the case of split rank 1 is clear in general.

The subalgebra $\cm \cong su(p-r) \oplus u(1) \oplus \cdots\oplus u(1)$, ($r$
factors), and is explicitly given as ($\b = \b_2$): \eqn\maxcm{
\eqalign{\cm
=  \{\ X = \pmatrix{ u_{p-r} & 0 & 0 \cr 0 & u & 0 \cr 0 & 0 & u} ~|~&
u_{p-r}^\dag = - u_{p-r}\ , ~~~u = i\, {\rm diag}\,
(\phi_1,\ldots,\phi_n), ~\phi_j\in\bbr \ ,\cr \qquad
&{\rm tr}\, u_{p-r} + 2{\rm tr}\, u = 0\
\}\ } } The subalgebras ~$\cn^\pm$~ which form the root spaces of
the root system $(\cg,\ca)$ are of real dimension ~$r(2p-1)$. The
restricted simple root system $(\cg,\ca)$ looks as that of the orthogonal
algebra $B_r$, however, the root spaces of the long roots have
multiplicity 2, the short simple root, say $\a_r\,$, has multiplicity $2(p-r)$,
and there is also a root $2\a_r$ with multiplicity 1.

Similarly to the $su(n,n)$ case one can prove that the
nontrivial cuspidal parabolic subalgebras are given by $\Th$ of the form:
\eqn\thitc{ \Th_j ~=~ \{\ j+1, \ldots, r\ \}\ , ~~~1 \leq j < r \ ,~~~r>1.}
The corresponding cuspidal parabolic subalgebras  contain the subalgebras
\eqn\cmmu{ \cm_{\Th_j} ~\cong~ su(p-j,r-j)\oplus  u(1)
\oplus \cdots\oplus  u(1), ~~~j~ {\rm factors} \ . }

The maximal parabolic subalgebras, (cf. \thitb), contain the   $\cm_\Th$
subalgebras are of the form:
\eqn\cmmax{ \cm^{\rm max}_j ~=~ sl(j,\bbc) \oplus su(p-j,r-j)\oplus  u(1)
\oplus \cdots\oplus  u(1), ~~~j~ {\rm factors} \ .  }
Thus, the only cuspidal maximal parabolic subalgebra is ~$\cp_{\Th_1}\,$.\nl
The latter is also the only Heisenberg parabolic subalgebra. %25.9.2021

\newsec{BDI,BDII :  $SO(p,r)$}

\nt
In this section $G=SO(p,r)$, $p\geq r$, which standardly is defined as
follows: \eqn\SOpr{SO(p,r) \doteq \{\ g\in SO(p+r,\bbc) ~|~ g^\dag
\b_0 g = \b_0 \ , ~~~\b_0 \equiv \pmatrix{1_p & 0 \cr 0 & -1_r
} \  \} \ ,} where $g^\dag$ is the Hermitian
conjugate of $g$.   We shall use also another
realization of $G$ differing from \SOpr{} by unitary transformation:
\eqn\SOppr{\b_0 \mt \b_2 \equiv \pmatrix{1_{p-r}& 0 & 0\cr 0 & 0 & 1_r \cr 0& 1_r & 0 } =
U\b_0 U^{-1} \ , \quad U \equiv {1\over\sqrt{2}} \pmatrix{1_{p-r} & 0 & 0\cr 0 & 1_r & 1_r
\cr 0& -1_r & 1_r} \ .} The Lie algebra $\cg=so(p,r)$ is given by
($\b=\b_0,\b_2$): \eqn\sopr{ so(p,r) \doteq \{\ X\in so(p+r,\bbc) ~|~
X^\dag \b + \b\,X = 0 \ \}\ . } The Cartan
involution is given explicitly by: ~$\th X = \b X\b$. Thus, $\ck
\cong so(p)\ \oplus so(r)$, and more explicitly is
given as ($\b = \b_0$): \eqn\maxcs{ \ck =  \{\ X = \pmatrix{u_1 &
0 \cr 0 & u_2} ~|~ u_1 \in so(p), ~~ u_2 \in so(r) \ \}\ . }

Note that ~$so(5,1) \cong su^*(4)$, ~$so(4,2) \cong su(2,2)$,
~$so(3,3) \cong sl(4,\bbr)$, ~$so(4,1) \cong sp(1,1)$,
~$so(3,2) \cong sp(2,\bbr)$, ~$so(3,1) \cong sl(2,\bbc)$,
~$so(2,2) \cong sl(2,\bbr)\oplus sl(2,\bbr)$,
~$so(2,1) \cong sl(2,\bbr)$, ($so(1,1)$ is abelian).
Thus, below we can restrict to ~$p+r >4$, since the cases $p+r=5$ are
not treated yet.

Note that $so(p,r)$  has discrete series representations
except when both $p,r$ are odd numbers, since then
~$\rank\, \ck = \rank\, so(p) + \rank\, so(r) = \ha (p+r-2)  < \rank\, so(p,r) = \ha (p+r)$.
It has highest/lowest weight representations when $p\geq r=2$ and $p=2,r=1$.

The split rank is equal to ~$r$~ and the abelian subalgebra $\ca$ may
be given explicitly by ($\b = \b_2$): \eqn\aao{\ca = {\rm r.l.s.} \{\
H^u_j   \equiv e_{p-r+j,p-r+j} - e_{p+j,p+j} \ , \quad 1\leq j \leq r\ \}\ .
}

The subalgebra $\cm \cong so(p-r)$ and is explicitly given as ($\b = \b_2$):
\eqn\masocm{ \cm
=  \{\ X = \pmatrix{ u & 0  \cr 0 &   0  } ~|~ u \in so(p-r)
\}\ }

The subalgebras ~$\cn^\pm$~ which form the root spaces of
the root system $(\cg,\ca)$ are of real dimension ~$r(p-1)$.
Except in the case ~$p=r$~ the restricted
simple root system $(\cg,\ca)$ looks as that of the orthogonal
algebra $B_r$, however, the short simple root, say $\a_r\,$,
has multiplicity $p-r$.

Thus, we consider first the case ~$p> r>1$.
First we note that the parabolic subalgebras given by $\Th_j = \{\ j\ \}, j<r$
contain a factor: ~$\cm_{\Th_j} = sl(2,\bbr) \oplus so(p-r)$. More generally,
if $r\notin \Th$ then all possible cuspidal parabolic subalgebras are like those of
~$sl(r,\bbr)$, adding the compact subalgebra ~$so(p-r)$.
Suppose now, that $r\in \Th$. In that case, $\Th$ will
include a set $\Th_j$ of the form:
\eqn\thitd{ \Th_j ~=~ \{\ j+1, \ldots, r\ \}\ , ~~~1 \leq j < r \ .}
That would bring a $\cm_\Th$ factor of the form ~$so(p-j,r-j)$.
Thus, all possible cuspidal parabolic subalgebras are obtained for those $j$,
for which the number $(p-j)(r-j)$ is even and for fixed such $j$ they
would be like those of
~$sl(j,\bbr)$, adding the non-compact subalgebra ~$so(p-j,r-j)$.
Clearly, if both $p,r$ are even (odd), then also $j$ must be even (odd),
while if one of $p,r$ is even and the other odd, i.e., $p+r$ is odd,
then $j$ takes all values from \thitd{}.

To be more explicit we first introduce the notation:
\eqn\seqv{ {\bar n}_s ~\doteq~ \{\ n_1,\ldots,n_s\ \}\ , ~~~ 1\leq s \leq r \ ,}
(note that ~${\bar n}_r = {\bar n}$~ from \sequ{}).
Then we shall use the notation introduced for the ~$sl(n,\bbr)$~ case,
namely, ~$\Th({\bar n}_s)$~ from \thitz{}. Then the cuspidal parabolic subalgebras are
given by the noncompact factors ~$\cm_\Th$~ from \cuspslnr{}:
\eqn\cuspsopr{ \cm_s ~=~ \cm_{\Th({\bar n}_s)}\oplus so(p-s,r-s)\ ,
\quad \cases{ s=1,2,\ldots,r-1 ~&~ $p+r$~odd \cr
s=2,4,\ldots,r-2 ~&~ $p,r$~even \cr
s=1,3,\ldots,r-2 ~&~ $p,r$~odd \cr
 }}

Next we note that we can include the case when the second factor in $\cm_\Th$ is compact
 by just extending the range of $s$ to $r$.

 Thus, all cuspidal parabolic subalgebras of ~$so(p,r)$~ in the case ~$p>r$~
 will be determined by the following $\cm_\Th$ subalgebras:
\eqn\cuspsoprf{ \cm_s ~=~ \cm_{\Th({\bar n}_s)} \oplus so(p-s,r-s)\ ,
\quad \cases{ s=1,2,\ldots,r ~&~ $p+r$~odd \cr
s=2,4,\ldots,r ~&~ $p,r$~even \cr
s=1,3,\ldots,r ~&~ $p,r$~odd \cr
 }}
The algebras ~$\cm_\Th$~ have highest/lowest weight representations
only when ~$s=r-2$~ or ~$s=r$, since then the second factor is
~$so(p-r+2,2)$, $so(p-r)$, resp.

Finally, we note that the maximal parabolic subalgebras corresponding to \thitb{}
have $\cm_\Th$-factors given by:
\eqn\cmsosmax{ \cm^{\rm max}_j ~=~ sl(j,\bbr) \oplus so(p-j,r-j) \ , \quad j=1,2,\ldots,r\ . }
Thus, the maximal parabolic subalgebras are cuspidal (and can be found in \cuspsoprf{})
when $j=1,2$ and the number $(p-j)(r-j)$ is even. In addition, ~$\cm^{\rm max}_j$~
 have highest/lowest weight representations only when ~$r-j=0,2$, (or
 $p-j=2$).\nl
 The  case $j=2$ is   the only Heisenberg parabolic subalgebra. %25.9.2021

Now we consider the split cases ~$p=r\geq 4$. (Note
that the other split-real cases, i.e., when $p=r+1$, were considered above
without any peculiarities. The split cases $p=r<4$ are not representative
of the situation and were treated already:
~$so(3,3) \cong sl(4,\bbr)$, ~ $so(2,2)\cong so(2,1) \oplus so(2,1)$, $so(1,1)$ is not semisimple.)
We accept the convention that the simple roots $\a_{r-1}$ and $\a_r$ form the fork of the $so(2r,\bbc)$
simple root system, while $\a_{r-2}$ is the simple root connected to the
simple roots $\a_{r-3}\,$, $\a_{r-1}$ and $\a_r\,$.
Special care is needed only when $\Th$ includes these four special roots, i.e.,
\eqn\thite{\Th ~\supset~ {\hat \Th}_s ~\doteq~ \{\ s, \ldots, r\ \}\ , ~~~1 \leq s \leq r-4 \ .}
In these cases, we have $\cm$ factor of the form ~$so(r-s,r-s)$, i.e.,
there will be no cuspidal parabolic if  ~$r-s$~ is odd.

For all other $\Th$ the parabolic subalgebras would be like those of
~$sl(r,\bbr)$, when $r\notin\Th$ or ${r-1}\notin\Th$, or like those
of ~$sl(r-2,\bbr)$~ with possible addition of one or two
$sl(2,\bbr)$ factors, (when ${r-2}\notin\Th$). To describe the
latter cases we need a modification of the notation \seqv{}:
\eqn\thity{ \Th^o({\bar n}) ~=~ \{\ j ~|~ n_j = 1 , ~~n_{j+1}=0~
{\rm if}~ j\neq r-1\ \}\ .}

Thus, the cuspidal parabolic subalgebras are determined by the
following $\cm_\Th$ factors:
\eqn\cuspsoprff{ \cm_\Th ~=~\cases{  \cm_{\Th({\bar n}_s)}\oplus so(r-s,r-s)\ ,
~&~ $\Th \supset{\hat \Th}_s\ ,\quad  \cases{s=2,4,\ldots,r-4 ~&~ $r$~even \cr
s=1,3,\ldots,r-4 ~&~ $r$~odd \cr}$ \cr &\cr
\cm_{\Th^o({\bar n})} \ , ~&~ $\Th \supset {\kern -10pt /}\ ~{\hat \Th}_s$  }}
Only the second subcase, namely, ~$\cm_{\Th^o({\bar n})}\,$, has highest/lowest weight representations.

The maximal parabolic subalgebras corresponding to \thitb{}
have $\cm_\Th$-factors given by:
\eqn\cmsosemax{ \cm^{\rm max}_j ~=~ \cases{
sl(r,\bbr) ~&~ $j=r-1,r$ \cr
sl(r-2,\bbr) \oplus sl(2,\bbr) \oplus sl(2,\bbr) ~&~ $j=r-2$ \cr
sl(r-3,\bbr) \oplus sl(4,\bbr)  ~&~ $j=r-3$ \cr
sl(j,\bbr) \oplus so(r-j,r-j)  ~&~ $j\leq r-4$ \cr}}
Thus the maximal parabolic subalgebras which are cuspidal occur for
$j=1$ and odd $r\geq 5$, (4th case), or
$j=2$ and either $r=4$,
(2nd case), or even $r\geq 6$, (4th case):
\eqn\cmsosemaxcusp{\eqalign{
& \cm^{\rm max}_1 ~=~ so(r-1,r-1)\ , \qquad r = 5,7,\ldots \cr
& \cm^{\rm max}_2 ~=~ \cases{
sl(2,\bbr) \oplus sl(2,\bbr) \oplus sl(2,\bbr) ~&~ $r=4$ \cr
sl(2,\bbr) \oplus so(r-2,r-2)  ~&~ $r= 6,8,\ldots$ \cr}} }
Of these, only ~$\cm^{\rm max}_2$~ for $r=4$ has   highest/lowest weight representations
(it belongs to the second subcase of \cuspsoprff).\nl
The  cases ~$\cm^{\rm max}_2$~ are  from the   Heisenberg parabolic subalgebras
(recall that ~$so(2,2) = sl(2,\bbr) \oplus sl(2,\bbr)$). %25.9.2021

\newsec{CI :  $Sp(n,\bbr),$ $n>1$}

\nt In this section $G=Sp(n,\bbr)$ - the split real form of
$Sp(n,\bbc)$. Both are standardly  defined by: \eqn\Spnc{Sp(n,F)
\doteq \{\ g\in GL(2n,F) ~|~\ ^tg J_n g = J_n  \ , ~~~\det\,g =1\ \}
\ , \qquad F = \bbr,\bbc \ . } Correspondingly, the Lie algebras are
given by: \eqn\spnc{ sp(n,F) = \{X\in gl(2n,F) ~|~\ ^tXJ_n + J_nX
=0\}\ . } Note that ~$\dim_F\,sp(n,F) ~=~ n(2n+1)$. The general
expression for $X \in sp(n,F)$ is \eqn\spnce{ X = \pmatrix{A & B \cr
C & -~^tA\cr}, \ \ A,B,C \in gl(n,F), \ \ ^tB = B, \ \ ^tC = C ~.} A
basis of the Cartan subalgebra ~$\ch$~ of $sp(n,\bbc)$ is:
\eqn\spca{\eqalign{ H_i =& \pmatrix{ A_i & 0\cr 0 & -A_i}, \ \ i =
1, \dots , n - 1, \qquad A_i = {\rm diag} (0, \dots 0,1, -1, 0,
\dots , 0), \cr H_n =& \pmatrix{ A'_n & 0\cr 0 & -A'_n},\ \qquad
A'_n = (0, \dots, 0,2). }} The same basis over ~$\bbr$~ spans the
subalgebra ~$\ca$~ of ~$\cg=sp(n,\bbr)$, since ~$\rank_F\,sp(n,F)
~=~ n$. Note that ~$sp(2,\bbr) \cong so(3,2)$, ~$sp(1,\bbr) \cong
sl(2,\bbr)$.

The maximal compact subalgebra of ~$\cg = sp(n,\bbr)$~ is ~$\ck
\cong u(n)$, thus ~$sp(n,\bbr)$~ has discrete series representations
(and highest/lowest weight representations). Explicitly, \eqn\spnce{
\ck = \{\ X = \pmatrix{A & 0 \cr 0 & -\ ^tA \cr} ~|~  A \in u(n) \
\} ~.}

The subalgebras ~$\cn^\pm$~ which form the root spaces of
the root system $(\cg,\ca)$ are of real dimension ~$n^2$.

Further, we choose the long root of the $C_n$ simple root system to be $\a_n\,$.

The parabolic subalgebras corresponding to $\Th$ such that $n\notin \Th$ are
the same as the parabolic subalgebras of $sl(n,\bbr)$.
The parabolic subalgebras corresponding to $\Th$ such that $n\in \Th$ contain a string
~$\Th'_s ~=~ \{\ s+1,\ldots,n\ \}$.
This string brings in $\cm_\Th$ a factor ~$sp(n-s,\bbr)$, which has
discrete series representations. Thus cuspidality depends on the rest of the
possible choices and are the same as the parabolic subalgebras of $sl(j,\bbr)$.
Thus, we have:
\eqn\thitspn{ \Th_s ~=~ \Th({\bar n}_{s-1}) \ \cup \ \Th'_s ,
\quad s=1,\ldots,n, }
with the convention that ~$\Th({\bar n}_{0})=\emptyset$,
~$\Th'_n =\emptyset$. Then the   $\cm_\Th$-factors
of the cuspidal parabolic subalgebras of ~$sp(n,\bbr)$~ are given as follows:
\eqn\cuspsp{
\cm_{\Th_s} ~=~  \cm_{\Th({\bar n}_{s-1})} \oplus sp(n-s,\bbr)\ , \quad s=1,\ldots,n\ . }

The minimal parabolic subalgebra for which ~$\cm_\Th=0$~ is obtained for ~$s=n$~
since then ~$\cm_{\Th({\bar n}_{n-1})}$~ enumerates all cuspidal parabolic subalgebras
of $sl(n,\bbr)$, including the minimal case $\cm_\Th=0$.

The maximal parabolic subalgebras, cf. \thitb{}, $1\leq j \leq n$,
contain    $\cm_\Th$ subalgebras of the form:
\eqn\cmsmax{ \cm^{\rm max}_j ~=~ sl(j,\bbr) \oplus sp(n-j,\bbr)   \ ,}
i.e., the only maximal cuspidal are those for $j=1,2$.\nl
The  case $j=1$ is   the only Heisenberg parabolic subalgebra. %25.9.2021

\newsec{CII :  $Sp(p,r)$}

\nt
In this section ~$G=Sp(p,r)$, $p\geq r$, which standardly is defined as
follows: \eqn\Sppr{Sp(p,r) \doteq \{\ g\in Sp(p+r,\bbc) ~|~ g^\dag
\g_0 g = \g_0 \ \} \ ,  ~~~\g_0 = \pmatrix{\b_0 & 0 \cr 0 &\b_0} \ ,}
and correspondingly the Lie algebra  $\cg=sp(p,r)$ is given by
 \eqn\sppr{ sp(p,r) \doteq \{\ X\in sp(p+r,\bbc) ~|~\
X^\dag \g_0 + \g_0\,X = 0 \ \}\ . }
The Cartan involution is given explicitly by: ~$\th X = \g_0 X\g_0\,$. Thus, $\ck
\cong sp(p)\ \oplus sp(r)$, and ~$\cg$~ has discrete series
representations (but not highest/lowest weight representations).
More explicitly:
\eqn\maxcs{ \ck =  \{\ X = \pmatrix{
u_1 & 0 & 0 &0 \cr 0 &u_2 &0& 0 \cr
0&0 &-u_1^\dag & 0 \cr
0&0&0&-u_2^\dag} ~|~ u_1 \in sp(p), ~~ u_2 \in sp(r) \ \}\ . }

The split rank is equal to ~$r$~ and the abelian subalgebra $\ca$ may
be given explicitly by:
\eqn\aasp{\ca = {\rm r.l.s.} \{\
H^s_j   \equiv e_{p-r+j,p-r+j} - e_{p+j,p+j}
- e_{2p+j,2p+j} + e_{2p+r+j,2p+r+j} \ , \quad 1\leq j \leq r\ \}\ .
}
The subalgebras ~$\cn^\pm$~ which form the root spaces of
the root system $(\cg,\ca)$ are of real dimension ~$r(4p-1)$.

The subalgebra ~$\cm ~\cong~ sp(p-r) \oplus sp(1) \oplus \cdots
 \oplus sp(1)$, $r$ factors.

Here, just for a moment we distinguish the cases $p>r$ and $p=r$,
since the restricted root systems  are different.

When ~$p>r$~ the restricted simple root system $(\cg,\ca)$ looks as that of
$B_r$, however, the short simple root say, $\a_r\,$, has
multiplicity $4(p-r)$, the long roots  have multiplicity $4$.

When ~$p=r>1$~ the restricted simple root system $(\cg,\ca)$ looks as that of
$C_r$, however, the long root say, $\a_r\,$, has
multiplicity 3, the short  roots  have multiplicity $4$.
(We consider  $r>1$ since ~$sp(1,1) \cong so(4,1)$.)

In spite of these differences from now on we can consider the two subcases together,
i.e., we take ~$p\geq r$.

There are two types of parabolic subalgebras depending on whether ~$r\notin \Th$~
or ~$r\in \Th$.

Let ~$r\notin \Th$. Then the parabolic subalgebras are like those of $su^*(2n)$.
Let $\Th$ enumerate a connected string of restricted simple roots:
~$\Th = S_{ij} = \{\ i,\ldots,j\ \}$, where $1\leq i\leq j <r$.
Then the corresponding subalgebra $\cm_\Th$ is given by:
\eqn\cmssprr{\cm_{ij} = su^*(2(s+1)) \oplus sp(1)  \oplus\cdots \oplus sp(1) \ ,
~~r-s-1~{\rm factors} \ , ~~~s\equiv j-i+1 \ .}
In general $\Th$ consists of such strings, each string of length $s$  produces
a factor $su^*(2(s+1))$, the rest of $\cm_\Th$ consists of $sp(1)\cong su(2)$ factors.
All these parabolic subalgebras are not cuspidal.

Let ~$r\in \Th$~ and consider the various strings containing $r$~:
\eqn\thitd{ \Th_j ~=~ \{\ j+1, \ldots, r\ \}\ , ~~~1 \leq j < r \ .}
The corresponding factor in $\cm_\Th$ is given by algebra
~$sp(p-j,r-j)$~ which has discrete series representations.
If $\Th$ contains in addition some other string then it would bring some $su^*(2(s+1))$
factor and the corresponding  $\cm_\Th$ will not have discrete series representations.
Thus the nontrivial cuspidal parabolic subalgebras are given by $\Th_j$
from \thitd{} and the corresponding  $\cm_\Th$ is:
\eqn\cuspsprr{ \cm_j \cong sp(p-j,r-j) \oplus sp(1) \oplus \cdots
 \oplus sp(1), \qquad j~ {\rm factors} \ .}
All these ~$\cm_j$~ do not have highest/lowest weight representations.
The other factors in the cuspidal parabolic subalgebras
have   dimensions: ~$\dim\,\ca_j ~=~j$,
~$\dim\,\cn^\pm_j ~=~j(4p+4r-4j-1)$.
 Extending the range of ~$j$~ we include the minimal parabolic subalgebra for ~$j=r$~ and the case
 ~$\cm'=\cp=\cg$~ for ~$j=0$.

The maximal parabolic subalgebras corresponding to \thitb{}
have $\cm_\Th$-factors given by:
\eqn\cmsprrmax{ \cm^{\rm max}_j ~=~ su^*(2j) \oplus sp(p-j,r-j)\  , ~~~1 \leq j \leq r\ .}
The ~$\cn^\pm$~  factors in the maximal parabolic subalgebras
have dimensions:  ~$\dim\,(\cn^\pm_j)^{\rm max} ~=$ $j(4p+4r-6j+1)$.
The only cuspidal maximal parabolic subalgebra is ~$\cp_{\Th_1}\,$,
using $su^*(2) \cong su(2) \cong sp(1)$ and noting that \cuspsprr{}
and \cmsprrmax{} coincide for $j=1$.

\newsec{DIII :   $SO^*(2n)$}

\nt The group ~$G=SO^*(2n)$~ consists of all matrices in
$SO(2n,\bbc)$ which commute with a real skew-symmetric matrix times
the complex conjugation operator $C$~: \eqn\SOs{SO^*(2n) \doteq \{\
g\in SO(2n,\bbc) ~|~ J_n C g= g J_n C \} \ .} The Lie algebra
$\cg=so^*(2n)$ is given by: \eqn\sos{\eqalign{ so^*(2n) &\doteq \{\
X\in so(2n,\bbc) ~|~ J_n C X= X J_n C  \} \ = \cr &= \ \{\ X=
\pmatrix{a & b \cr - {\bar b} & {\bar a}} ~|~\ a,b\in gl(n,\bbc), ~~
^ta = -a, ~~ b^\dag = b \  \}\ .}} $\dim_R\,\cg = n(2n-1)$,
$\rank\,\cg =n$.

Note that ~$so^*(8) \cong so(6,2)$, ~$so^*(6) \cong su(3,1)$,
~$so^*(4) \cong so(3) \oplus so(2,1)$, ~$so^*(2) \cong so(2)$.
Further, we can restrict to ~$n\geq 4$~ since the other cases are not representative.

The Cartan involution is given   by: ~$\th X = -X^\dag$. Thus, $\ck
\cong u(n)$: \eqn\susc{\ck = \{\ X= \pmatrix{a & b \cr -b & a} ~|~\
 a,b\in gl(n,\bbc), ~~^ta = -a= -{\bar a}, ~~b^\dag = b = {\bar b}
\ \}\ , } and  $\cg = so^*(2n)$ has discrete series representations
(and highest/lowest weight representations).  The complimentary
space $\cp$ is given by: \eqn\susc{\cp = \{\ X= \pmatrix{a & b \cr b
& -a} ~|~\ a,b\in gl(n,\bbc) \ , ~~ ^ta = -a = {\bar a}, ~~b^\dag =
b = -{\bar b}\
 \ \}\ . }
$\dim_R\,\cp = n(n-1)$.
The split rank is  ~$r\equiv [n/2]$.

The subalgebras ~$\cn^\pm$~ which form the root spaces of
the root system $(\cg,\ca)$ are of real dimension ~$n(n-1) - [n/2]$.

Here, just for a moment we distinguish the cases $n$ even and $n$ odd
since the subalgebras ~$\cm$~ and the restricted root
systems are different.

For  ~$n=2r$~ the split rank is equal to $r\geq 2$, and
the restricted root system is as that of $C_r\,$,
but the short roots have multiplicity $4$, while
the long simple root $\a_r$ has multiplicity~1.
The subalgebra ~$\cm ~\cong~  so(3) \oplus \cdots
 \oplus so(3)$, $r$ factors.

For  ~$n=2r+1$~ the  split rank is equal to $r\geq 2$, and
the restricted root system is as that of $B_r\,$,
but all simple roots have multiplicity $4$,
and there is a restricted root ~$2\a_r$~ of multiplicity 1,
where  ~$\a_r$~ is the short simple root.\nl
The subalgebra ~$\cm ~\cong~ so(2) \oplus  so(3) \oplus \cdots
 \oplus so(3)$, $r$ factors.

In spite of these differences from now on we can consider the two subcases together.

There are two types of parabolic subalgebras depending on whether ~$r\notin \Th$~
or ~$r\in \Th$.

If ~$r\notin \Th$~ then the parabolic subalgebras are like those of $su^*(2r)$,
and they are not cuspidal.

Let ~$r\in \Th$~ and consider the various strings containing $r$~:
\eqn\thite{ \Th_j ~=~ \{\ j+1, \ldots, r\ \}\ , ~~~1 \leq j < r = [n/2]\ .}
The corresponding factor in $\cm_\Th$ is given by the algebra
~$so^*(2n - 4j)$~ which has discrete series representations. Thus, all
cuspidal parabolic subalgebras are enumerated by \thite{} and are:
\eqn\cuspsose{\cm_j ~=~ so^*(2n  -4j) \oplus so(3) \oplus \cdots
 \oplus so(3), \qquad j~ {\rm factors} \ , ~~~j=1,\ldots,r-1 \ .}
 All these ~$\cm_j$~ have highest/lowest weight representations.
 The other factors in the cuspidal parabolic subalgebras
have   dimensions: ~$\dim\,\ca_j ~=~j$,
~$\dim\,\cn^\pm_j ~=~j(4n-4j-3)$.
Extending the range of ~$j$~ we include  the minimal parabolic case for ~$j=r=[n/2]$, and
the case  ~$\cm'=\cp=\cg$~ for ~$j=0$~ which is also cuspidal.

The maximal parabolic subalgebras enumerated by $\Th^{\rm max}_j$
from \thitb{} have $\cm_\Th$-factors as follows:
\eqn\maxsose{ \cm^{\rm max}_j ~=~ so^*(2n-4j) \oplus su^*(2j)  \ , ~~~j=1,\ldots,r\ . }
The ~$\cn^\pm$~  factors in the maximal parabolic subalgebras
have dimensions:  ~$\dim\,(\cn^\pm_j)^{\rm max} ~=$ $j(4n-6j-1)$.
Only the case $j=1$ is cuspidal, noting that ~$\cm^{\rm max}_1$~
coincides with ~$\cm_1$~ from \cuspsose{}, ($su^*(2) \cong su(2) \cong so(3)$).\nl
The  case $j=1$ is  also the only Heisenberg parabolic subalgebra. %25.9.2021

\newsec{Real forms of the exceptional simple Lie algebras}

\nt
We start with the real forms of the exceptional simple Lie algebras.
Here we can not be so explicit with the matrix realizations.
To compensate this we use the Satake diagrams \Sat,\War, which we omitted
until now.\foot{We could do this, since by being explicit we could consider
simultaneously cases with different Satake diagrams.} A Satake diagram
has a starting point the Dynkin diagram of the corresponding complex form.
For a split real form it remains the same. In the other cases some dots
are painted in black - these considered by themselves are Dynkin diagrams
of the compact semisimple factors ~$\cm$~ of the minimal parabolic subalgebras.
Further, there are arrows connecting some nodes which use the $\bbz_2$ symmetry
of some Dynkin diagrams.  Then the reduced root systems are described by
Dynkin-Satake diagrams which are obtained from the Satake diagrams by dropping the black nodes,
identifying the arrow-related nodes, and adjoining all nodes in a connected Dynkin-like diagram,
but in addition noting the multiplicity of the
reduced roots (which is in general different from 1). More details can be seen in \War,
and we have tried to make the exposition transparent (by repeating things).

\newsubsec{EI :~   $E'_6$}

\nt
The split real form of ~$E_6$~ is denoted  as  ~$E'_6\,$, sometimes as ~$E_{6(+6)}\,$.
The maximal compact subgroup is ~$\ck \cong sp(4)$, ~$\dim_\bbr\,\cp = 42$,
~$\dim_\bbr\,\cn^\pm = 36$.
This real form does not have discrete series representations.

For a split real form the Satake diagram coincides with the Dynkin
diagram of the corresponding complex Lie algebra \War.

In the present case this Dynkin-Satake diagram is taken as follows:
\eqn\satsixa{ \downcirc{{\a_1}} \riga\downcirc{\a_2} \riga
\downcirc{{\a_3}}\kern-8pt\raise11pt\hbox{$\vert$}
\kern-3.5pt\raise22pt\hbox{$\circ {{\scriptstyle{\a_6}}}$}
\riga\downcirc{{\a_4}} \riga\downcirc{{\a_5}} }

Taking into account the above enumeration of simple roots
the cuspidal parabolic subalgebras have $\cm_\Th$-factors as follows:
\eqn\cuspsixa{
\cm_\Th ~=~
\cases{0 ~&~ $\Th = \emptyset$, ~minimal \cr
sl(2,\bbr)_j \ ,~&~ $\Th = \{j\}, ~~~j=1,\ldots, 6 $\cr
sl(2,\bbr)_j \oplus sl(2,\bbr)_k \ ,~&~ $\Th = \{j,k\}\ : ~ j+1<k,
~~~\{j,k\} \neq \{3,6\}; $\cr
&$\qquad (j,k) =(5,6),  $\cr
sl(2,\bbr)_j \oplus sl(2,\bbr)_k \oplus sl(2,\bbr)_\ell\ ,
&~ $\Th = \{j,k,\ell\}, ~~~(j,k,\ell) = (1,3,5),(1,4,6),$\cr
&$\qquad (1,5,6),(2,4,6),(2,5,6)$\cr
so(4,4) \ ,~&~ $\Th = \{2,3,4,6\}$, \cr
}}
where ~$sl(2,\bbr)_j$~ denotes the ~$sl(2,\bbr)$~
subalgebra of $\cg$ spanned  by $X^\pm_j\,, H_j\,$,
(using the same notation as in the Section on $sl(n,\bbr)$).
All these ~$\cm_\Th\,$, except the last case ($so(4,4)$), have highest/lowest weight representations.

The dimensions of the other factors are, respectively:
\eqn\cuspsixab{
\dim\,\ca_\Th ~=~
\cases{ 6 \cr 5 \cr  4 \cr 3 \cr 2 \cr }\ , \qquad
\dim\,\cn^\pm_\Th ~=~
\cases{ 36 \cr 35 \cr  34 \cr 33 \cr 24 \cr }
}

Taking into account \thitb{}
the maximal parabolic subalgebras are determined by:
\eqn\maxsixa{\eqalign{
& \cm^{\rm max}_1 \cong \cm^{\rm max}_5 \cong so(5,5)
\ , \qquad \dim\,(\cn^\pm_\Th)^{\rm max} ~=~ 16
\cr
& \cm^{\rm max}_2 \cong \cm^{\rm max}_5 \cong
sl(5,\bbr) \oplus sl(2,\bbr)
\ , \qquad \dim\,(\cn^\pm_\Th)^{\rm max} ~=~ 25
\cr
& \cm^{\rm max}_3 \cong
sl(3,\bbr) \oplus sl(3,\bbr) \oplus sl(2,\bbr)
\ , \qquad \dim\,(\cn^\pm_\Th)^{\rm max} ~=~ 29
\cr
& \cm^{\rm max}_6 \cong sl(6,\bbr)
\ , \qquad \dim\,(\cn^\pm_\Th)^{\rm max} ~=~ 21
 }}
Clearly, no maximal parabolic subalgebra is cuspidal.
The  case $\cm^{\rm max}_6$ is   the only Heisenberg parabolic subalgebra. %25.9.2021

\newsubsec{EII :~   $E''_6$}

\nt
Another real form of ~$E_6$~ is denoted as ~$E''_6\,$, sometimes as ~$E_{6(+2)}\,$.
The maximal compact subgroup is ~$\ck \cong su(6)\oplus su(2)$, ~$\dim_\bbr\,\cp = 40$,
~$\dim_\bbr\,\cn^\pm = 36$.
This real form has discrete series representations.

The split rank is equal to 4, while ~$\cm \cong u(1)\oplus u(1)$.

The Satake diagram is:
\eqn\satsixaz{\underbrace{ \downcirc{{\a_1}} \riga \underbrace{ \downcirc{\a_2} \riga
\downcirc{{\a_3}}\kern-8pt\raise11pt\hbox{$\vert$}
\kern-3.5pt\raise22pt\hbox{$\circ {{\scriptstyle{\a_6}}}$}
\riga\downcirc{{\a_4}}} \riga\downcirc{{\a_5}} }}
Thus, the reduced root system is presented
by a Dynkin-Satake diagram looking like the ~$F_4$~ Dynkin diagram:
\eqn\satsixb{ \downcirc{{\l_1}}
\riga\downcirc{{\l_2}} \Longrightarrow\downcirc{{\l_3}}
\riga\downcirc{{\l_4}}}
but the short roots have multiplicity $2$
(the long - multiplicity $1$).
It is obtained from \satsixaz{} by identifying
 ~$\a_1$~ and ~$\a_5$~ and mapping them to ~$\l_4$,
identifying
 ~$\a_2$~ and ~$\a_4$~ and mapping them to ~$\l_3$,
while the roots ~$\a_3,\a_6$~ are mapped to
the $F_4$-like long simple roots ~$\l_2,\l_1\,$, resp.

Using the above enumeration of $F_4$ simple roots we give the $\cm_\Th$-factors of all
parabolic subalgebras:
\eqn\parabsixb{
\cm_\Th ~=~ \cases{
u(1)\oplus u(1)\ ,~&~ $\Th = \emptyset$, minimal \cr
sl(2,\bbr)_j \oplus u(1)\oplus u(1)\ ,~&~ $\Th = \{j\}, ~~~j=1,2$\cr
sl(2,\bbc)_j \oplus u(1)\ ,~&~ $\Th = \{j\}, ~~~j=3,4$\cr
sl(3,\bbr) \oplus u(1)\oplus u(1)\ ,~&~ $\Th = \{1,2\}$\cr
sl(2,\bbr)_1 \oplus sl(2,\bbc)_j  \oplus u(1)\ ,~&~ $\Th = \{1,j\}, ~~~j=3,4$\cr
sl(4,\bbr) \oplus u(1)\ ,~&~ $\Th = \{2,3\}$\cr
sl(2,\bbr)_2 \oplus sl(2,\bbc)_4  \oplus u(1)\ ,~&~ $\Th = \{2,4\}$\cr
sl(3,\bbc) \ ,~&~ $\Th = \{3,4\}$\cr
so(5,3) \oplus u(1) \ ,~&~ $\Th = \{1,2,3\}$, \cr
sl(3,\bbr) \oplus u(1)\oplus sl(2,\bbc)_4 \ ,~&~ $\Th = \{1,2,4\}$\cr
sl(2,\bbr)_1\oplus sl(3,\bbc) \ ,~&~ $\Th = \{1,3,4\}$\cr
su(3,3)\ ,~&~ $\Th = \{2,3,4\}$\cr
}}
The dimensions of the other factors are, respectively:
\eqn\cuspsixbb{
\dim\,\ca_\Th ~=~
\cases{ 4 \cr 3 \cr  3 \cr 2 \cr 2 \cr 2 \cr 2 \cr  2 \cr 1 \cr 1 \cr  1\cr 1}\ , \qquad
\dim\,\cn^\pm_\Th ~=~
\cases{ 36 \cr 35 \cr  34 \cr 33 \cr 33 \cr 30 \cr 33 \cr  30 \cr 24 \cr 31 \cr 29 \cr 21 \cr}
}

The maximal parabolic subalgebras are given the last four lines in the above lists
corresponding to ~$\Th^{\rm max}_j\,$, $j=4,3,2,1$, cf. \thitb{}.\nl
The last case with $su(3,3)$ is   the only Heisenberg parabolic subalgebra. %25.9.2021

The cuspidal parabolic subalgebras are those containing:
\eqn\cuspsixb{
\cm_\Th ~=~ \cases{
u(1)\oplus u(1)\ ,~&~ $\Th = \emptyset$, minimal \cr
sl(2,\bbr)_j \oplus u(1)\oplus u(1)\ ,~&~ $\Th = \{j\}, ~~~j=1,2$\cr
su(3,3) \oplus u(1) \ ,~&~ $\Th = \{1,2,3\}$ \cr}}
All these ~$\cm_\Th\,$, except the last case (containing $so(4,4)$), have highest/lowest weight representations.
The last case $\cm^{\rm max}_6$ is   the only Heisenberg parabolic subalgebra. %25.9.2021

\newsubsec{EIII :~   $E'''_6$}

\nt
Another real form of ~$E_6$~ is denoted as $E'''_6\,$, sometimes as ~$E_{6(-14)}\,$.
The maximal compact subgroup is ~$\ck \cong so(10)\oplus so(2)$, ~$\dim_\bbr\,\cp = 32$,
~$\dim_\bbr\,\cn^\pm = 30$.
This real form has discrete series representations (and highest/lowest weight
representations).

The split rank is equal to 2, while ~$\cm \cong so(6)\oplus so(2)$.

The Satake diagram is:
\eqn\satsixay{\underbrace{ \downcirc{{\a_1}} \riga  \black{\a_2} \riga
\black{{\a_3}}\kern-8pt\raise11pt\hbox{$\vert$}
\kern-3.5pt\raise22pt\hbox{$\circ {{\scriptstyle{\a_6}}}$}
\riga\black{{\a_4}} \riga\downcirc{{\a_5}} }}
Thus, the reduced root system is presented
by a Dynkin-Satake diagram looking like the ~$B_2$~ Dynkin diagram
but the long roots (incl. $\l_1$) have multiplicity 6, while the short roots
(incl. $\l_2$) have multiplicity 8, and there are also the roots $2\l$
of multiplicity 1, where $\l$ is any short root.
It is obtained from \satsixay{} by dropping the black nodes,
(they give rise to $\cm$),
identifying ~$\a_1$~ and ~$\a_5$~ and mapping them to ~$\l_2$,
while the root ~$\a_6$~ is mapped to the long simple root ~$\l_1\,$.

The non-minimal   parabolic subalgebras are given by:
\eqn\parabsixc{
\cm_\Th ~=~ \cases{
so(7,1) \oplus so(2) \ ,~&~ $\Th = \{1\}$\cr
su(5,1)  \ ,~&~ $\Th = \{2\}$\cr}\ ,\qquad
\dim\,\cn^\pm_\Th ~=~
\cases{ 24 \cr 21  \cr}}
Both are maximal ($\dim\,\ca_\Th =1$), the second is cuspidal.\nl
The last case  is also    the only Heisenberg parabolic subalgebra. %25.9.2021

\newsubsec{EIV :~   $E^{iv}_6$}

\nt
Another real form of ~$E_6$~ is denoted as ~$E^{iv}_6\,$, sometimes as ~$E_{6(-26)}\,$.
The maximal compact subgroup is ~$\ck \cong f_4$, ~$\dim_\bbr\,\cp = 26$,
~$\dim_\bbr\,\cn^\pm = 24$.
This real form does not have discrete series representations.

The split rank is equal to 2, while ~$\cm \cong so(8)$.

The Satake diagram is:
\eqn\satsixd{
\downcirc{ {\a_1}} \riga \black{\a_2} \riga
\black{{\a_3}}\kern-8pt\raise11pt\hbox{$\vert$}
\kern-3.5pt\raise22pt\hbox{$\bullet
{{\scriptstyle{\a_6}}}$} \riga\black{{\a_4}}
\riga\downcirc{{\a_5}}  }
Thus, the reduced root system is presented
by a Dynkin-Satake diagram looking like the ~$A_2$~ Dynkin diagram
but all roots   have multiplicity~8.
It is obtained from \satsixd{} by dropping the black nodes,
 while $\a_1\,,\a_5\,$, resp.,
are mapped to the $A_2$-like simple roots $\l_1\,,\l_2\,$.

The two non-minimal   parabolic subalgebras are isomorphic and given by:
\eqn\parabsixd{
\cm_\Th ~=~ so(9,1)  \ ,\quad \Th = \{j\}, ~~~j=1,2\ , \qquad
\dim\,\cn^\pm_\Th ~=~ 16 \ .}
Both are maximal ($\dim\,\ca_\Th =1$) and not cuspidal.

\newsubsec{EV :~   $E'_7$}

\nt
The split real form of ~$E_7$~ is denoted as ~$E'_7\,$, sometimes as ~$E_{7(+7)}\,$.
The maximal compact subgroup ~$\ck \cong su(8)$, ~$\dim_\bbr\,\cp = 70$,
~$\dim_\bbr\,\cn^\pm = 63$.
This real form has discrete series representations.

We take the Dynkin-Satake diagram as follows:
\eqn\satseva{
\downcirc{{\a_1}} \riga\downcirc{\a_2} \riga
\downcirc{{\a_3}}\kern-8pt\raise11pt\hbox{$\vert$}
\kern-3.5pt\raise22pt\hbox{$\circ
{{\scriptstyle{\a_7}}}$} \riga\downcirc{{\a_4}}
\riga\downcirc{{\a_5}} \riga\downcirc{{\a_6}}}

Taking into account the above enumeration of simple roots
the cuspidal parabolic subalgebras have $\cm_\Th$-factors as follows:
\eqn\cuspseva{
\cm_\Th ~=~
\cases{
0 ~&~ $\Th = \emptyset$, ~minimal \cr
sl(2,\bbr)_j \ ,~&~ $\Th = \{j\}, ~~~j=1,\ldots, 7 $\cr
sl(2,\bbr)_j \oplus sl(2,\bbr)_k\ ,
~&~ $\Th = \{j,k\}\ : ~ j+1<k,$ \cr
&$~\{j,k\} \neq \{3,7\}; ~~ \{j,k\} = \{6,7\},  $ \cr
sl(2,\bbr)_j \oplus sl(2,\bbr)_k \oplus sl(2,\bbr)_\ell\ ,
&$\Th = \{j,k,\ell \} =\{1,3,5 \},  \{1,3,6\},$\cr
&$\{1,4,6\},\{1,4,7\},\{1,5,7\},\{1,6,7\},$\cr
&$\{2,4,6\},\{2,4,7\},\{2,5,7\},\{2,6,7\}$\cr
sl(2,\bbr)_1 \oplus sl(2,\bbr)_4 \oplus sl(2,\bbr)_6 \oplus sl(2,\bbr)_7 \cr
sl(2,\bbr)_2 \oplus sl(2,\bbr)_4 \oplus sl(2,\bbr)_6 \oplus sl(2,\bbr)_7 \cr
so(4,4) \ ,~&~ $\Th = \{2,3,4,7\}$, \cr
so(6,6) \ ,~&~ $\Th = \{2,3,4,5,6,7\}$, \cr
}}
All these ~$\cm_\Th\,$, except the last two cases ($so(4,4),so(6,6)$), have highest/lowest weight representations.

The dimensions of the other factors are, respectively:
\eqn\cuspsevab{
\dim\,\ca_\Th ~=~
\cases{ 7 \cr 6 \cr  5 \cr 4 \cr 3 \cr 3 \cr 3 \cr 1}\ , \qquad
\dim\,\cn^\pm_\Th ~=~
\cases{ 63 \cr 62 \cr  61 \cr 60 \cr 59 \cr 59 \cr 51 \cr  33 \cr}
}

Taking into account \thitb{}
the maximal parabolic subalgebras are determined by:
\eqn\maxsixa{\eqalign{
& \cm^{\rm max}_1   \cong so(6,6)
\ , \qquad \dim\,(\cn^\pm_\Th)^{\rm max} ~=~ 33
\cr
& \cm^{\rm max}_2 \cong sl(6,\bbr) \oplus sl(2,\bbr)
\ , \qquad \dim\,(\cn^\pm_\Th)^{\rm max} ~=~ 47
\cr
& \cm^{\rm max}_3 \cong sl(4,\bbr) \oplus sl(3,\bbr) \oplus sl(2,\bbr)
\ , \qquad \dim\,(\cn^\pm_\Th)^{\rm max} ~=~ 53
\cr
&\cm^{\rm max}_4 \cong sl(5,\bbr) \oplus sl(3,\bbr)
\ , \qquad \dim\,(\cn^\pm_\Th)^{\rm max} ~=~ 60
\cr
& \cm^{\rm max}_5   \cong so(5,5) \oplus sl(2,\bbr)
\ , \qquad \dim\,(\cn^\pm_\Th)^{\rm max} ~=~ 42
\cr
& \cm^{\rm max}_6 \cong E'_6
\ , \qquad \dim\,(\cn^\pm_\Th)^{\rm max} ~=~ 27
\cr
&\cm^{\rm max}_7 \cong sl(7,\bbr)
\ , \qquad \dim\,(\cn^\pm_\Th)^{\rm max} ~=~ 42
\cr
}}
Clearly, the only  maximal cuspidal parabolic subalgebra is
the one containing ~$\cm^{\rm max}_1\,$.\nl
The latter case  is also    the only Heisenberg parabolic subalgebra. %25.9.2021

\newsubsec{EVI :~   $E''_7$}

\nt
Another real form of ~$E_7$~ is denoted as $E''_7\,$, sometimes as ~$E_{7(-5)}\,$.
The maximal compact subgroup is ~$\ck \cong so(12)\oplus su(2)$, ~$\dim_\bbr\,\cp = 64$,
~$\dim_\bbr\,\cn^\pm = 60$.
This real form has discrete series representations.

The split rank is equal to 4, while ~$\cm \cong su(2)\oplus su(2) \oplus su(2)$.

The Satake diagram is:
\eqn\satsevb{
\downcirc{{\a_1}} \riga\downcirc{\a_2} \riga
\downcirc{{\a_3}}\kern-8pt\raise11pt\hbox{$\vert$}
\kern-3.5pt\raise22pt\hbox{$\bullet
{{\scriptstyle{\a_7}}}$} \riga\black{{\a_4}}
\riga\downcirc{{\a_5}} \riga\black{{\a_6}}}
Thus, the reduced root system is presented
by a Dynkin-Satake diagram looking like the ~$F_4$~ Dynkin diagram,
cf. \satsixb, but the short roots have multiplicity $4$
(the long - multiplicity $1$).
Going to this  Dynkin-Satake diagram we drop the black nodes,
(they give rise to $\cm$), while $\a_1\,,\a_2\,,\a_3\,,\a_5\,$,
are mapped to $\l_1\,,\l_2\,,\l_3\,,\l_4$, resp., of \satsixb.

Using the above enumeration of $F_4$ simple roots
we shall give the $\cm_\Th$-factors of all
parabolic subalgebras:
\eqn\parabsevb{
\cm_\Th ~=~ \cases{
\cm = su(2)\oplus su(2) \oplus su(2) \ ,~&~ $\Th = \emptyset\ , ~~{\rm minimal}$ \cr
sl(2,\bbr)_j \oplus \cm\ ,~&~ $\Th = \{j\}, ~~~j=1,2$\cr
su^*(4) \oplus su(2)_{j+3}\ ,~&~ $\Th = \{j\}, ~~~j=3,4$\cr
sl(3,\bbr) \oplus \cm\ ,~&~ $\Th = \{1,2\}$\cr
sl(2,\bbr)_1 \oplus  su^*(4) \oplus su(2)_{j+3}\ ,~&~ $\Th = \{1,j\}, ~~~j=3,4$\cr
so(6,2) \oplus su(2)\ ,~&~ $\Th = \{2,3\}$\cr
sl(2,\bbr)_2 \oplus su^*(4) \oplus su(2)_{7}  \ ,~&~ $\Th = \{2,4\}$\cr
su^*(6) \ ,~&~ $\Th = \{3,4\}$\cr
so(7,3) \oplus su(2)_{6} \ ,~&~ $\Th = \{1,2,3\}$, \cr
sl(3,\bbr) \oplus su^*(4) \oplus su(2)_{7} \ ,~&~ $\Th = \{1,2,4\}$\cr
sl(2,\bbr)_1\oplus  su^*(6) \ ,~&~ $\Th = \{1,3,4\}$\cr
so^*(12)\ ,~&~ $\Th = \{2,3,4\}$\cr
}}
The dimensions of the other factors are, respectively:
\eqn\cuspsevab{
\dim\,\ca_\Th ~=~
\cases{ 4 \cr 3 \cr  3 \cr 2 \cr 2 \cr 2 \cr 2 \cr 2\cr 1 \cr 1 \cr 1 \cr 1
}\ , \qquad
\dim\,\cn^\pm_\Th ~=~
\cases{ 60 \cr 59 \cr  56 \cr 57 \cr 55 \cr 50 \cr 55 \cr 48 \cr 42
\cr 53\cr 47 \cr  33\cr
}}

The maximal parabolic subalgebras are the last four in the above list
corresponding to ~$\Th^{\rm max}_j\,$, $j=4,3,2,1$, cf. \thitb{}.

The cuspidal parabolic subalgebras are those containing
\eqn\cuspsevb{
\cm_\Th ~=~ \cases{
\cm = su(2)\oplus su(2) \oplus su(2) \ ,~&~ $\Th = \emptyset\ , ~~{\rm minimal}$ \cr
sl(2,\bbr)_j \oplus \cm\ ,~&~ $\Th = \{j\}, ~~~j=1,2$\cr
so(6,2) \oplus su(2)\ ,~&~ $\Th = \{2,3\}$\cr
so^*(12)\ ,~&~ $\Th = \{2,3,4\}$\cr
}}
the last one being also maximal.\nl  The last case  is also    the only Heisenberg parabolic subalgebra. %25.9.2021
\nl All these  ~$\cm_\Th$~ have highest/lowest weight representations.

\newsubsec{EVII :~   $E'''_7$}

\nt
Another real form of ~$E_7$~ is denoted as ~$E'''_7\,$, sometimes as ~$E_{7(-25)}\,$.
The maximal compact subgroup is ~$\ck \cong e_6\oplus so(2)$, ~$\dim_\bbr\,\cp = 54$,
~$\dim_\bbr\,\cn^\pm = 51$.
This real form has discrete series representations (and highest/lowest
weight representations).

The split rank is equal to 3, while ~$\cm \cong so(8)$.

The Satake diagram is:
\eqn\satsevb{
\downcirc{{\a_1}} \riga\black{\a_2} \riga
\black{{\a_3}}\kern-8pt\raise11pt\hbox{$\vert$}
\kern-3.5pt\raise22pt\hbox{$\bullet
{{\scriptstyle{\a_7}}}$} \riga\black{{\a_4}}
\riga\downcirc{{\a_5}} \riga\downcirc{{\a_6}}}
Thus, the reduced root system is presented
by a Dynkin-Satake diagram looking like the ~$C_3$~ Dynkin diagram:
\eqn\satsevc{
\downcirc{{\l_1}} \Longrightarrow\downcirc{{\l_2}} \riga\downcirc{{\l_3}}}
but the short roots have
multiplicity $8$  (the long - multiplicity $1$).
Going to the $C_3$ diagram we drop the black nodes,
(they give rise to $\cm$), while $\a_1\,,\a_5\,,\a_6\,$,
are mapped to $\l_3\,,\l_2\,,\l_1\,$, resp., of \satsevc.

Using the above enumeration of $C_3$ simple roots
we shall give the $\cm_\Th$-factors of all
parabolic subalgebras:
\eqn\parabsevb{
\cm_\Th ~=~ \cases{
  so(8) \ ,~&~ $\Th = \emptyset\ , ~~{\rm minimal}$ \cr
so(9,1)\ ,~&~ $\Th = \{j\}, ~~~j=1,2$\cr
sl(2,\bbr)_3 \oplus so(8)\ ,~&~ $\Th = \{3\}$\cr
e^{iv}_6\   ,~&~ $\Th = \{1,2\}$\cr
sl(2,\bbr)_3 \oplus so(9,1) \ ,~&~ $\Th = \{1,3\} $\cr
so(10,2)\ ,~&~ $\Th = \{2,3\}$\cr
}}
The dimensions of the other factors are, respectively:
\eqn\cuspsevab{
\dim\,\ca_\Th ~=~
\cases{ 3 \cr 2 \cr  2 \cr 1 \cr 1 \cr 1
}\ , \qquad
\dim\,\cn^\pm_\Th ~=~
\cases{ 51 \cr 43 \cr  50 \cr 27 \cr 42 \cr  33
}}
The last three give rise to the maximal parabolic subalgebras.

The cuspidal parabolic subalgebras are those containing
\eqn\cuspsevb{
\cm_\Th ~=~ \cases{
 so(8) \ ,~&~ $\Th = \emptyset\ , ~~{\rm minimal}$ \cr
sl(2,\bbr)_3 \oplus so(8)\ ,~&~ $\Th = \{3\}$\cr
so(10,2)\ ,~&~ $\Th = \{2,3\}$\cr}
}
the last one being also maximal.\nl  The last case  is also    the only Heisenberg parabolic subalgebra. %25.9.2021
\nl
All these   ~$\cm_\Th$~ have highest/lowest weight representations.

\newsubsec{EVIII :~   $E'_8$}

\nt
The split real form of ~$E_8$~ is denoted as ~ $E'_8\,$, sometimes as ~$E_{8(+8)}\,$.
The maximal compact subgroup ~$\ck \cong so(16)$, ~$\dim_\bbr\,\cp = 128$,
~$\dim_\bbr\,\cn^\pm = 120$.
This real form has discrete series representations.

We take the Dynkin-Satake diagram as follows:
\eqn\sateighta{
\downcirc{{\a_1}} \riga\downcirc{\a_2} \riga
\downcirc{{\a_3}}\kern-8pt\raise11pt\hbox{$\vert$}
\kern-3.5pt\raise22pt\hbox{$\circ
{{\scriptstyle{\a_8}}}$} \riga\downcirc{{\a_4}}
\riga\downcirc{{\a_5}} \riga\downcirc{{\a_6}}  \riga\downcirc{{\a_7}}  }

Taking into account the above enumeration of simple roots
the cuspidal parabolic subalgebras have $\cm_\Th$-factors as follows:
\eqn\cuspeighta{
\cm_\Th ~=~
\cases{
0 ~&~ $\Th = \emptyset$, ~minimal \cr
sl(2,\bbr)_j \ ,~&~ $\Th = \{j\}, ~~~j=1,\ldots, 8 $\cr
sl(2,\bbr)_j \oplus sl(2,\bbr)_k\ ,
~&~ $\Th = \{j,k\}\ : ~ j+1<k,$ \cr
&$~\{j,k\} \neq \{3,8\}; ~~ \{j,k\} = \{7,8\},  $ \cr
sl(2,\bbr)_j \oplus sl(2,\bbr)_k \oplus sl(2,\bbr)_\ell\ ,
&$\Th = \{j,k,\ell \} =\{1,3,5 \},  \{1,3,6\}, $\cr
&$ \{1,3,7\}, \{1,4,6\}, \{1,4,7\}, \{1,4,8\},$\cr
&$  \{1,5,7\}, \{1,5,8\},\{1,6,8\}, \{2,4,6\}, $\cr
&$ \{2,4,7\},  \{2,4,8\},\{2,5,7\}, \{2,5,8\},$\cr
&$  \{2,6,8\}, \{2,7,8\}, \{3,5,7\}, \{4,6,8\},$\cr
&$  \{4,7,8\}, \{5,7,8\}$\cr
sl(2,\bbr)_1 \oplus sl(2,\bbr)_3 \oplus sl(2,\bbr)_5 \oplus sl(2,\bbr)_7 \cr
sl(2,\bbr)_1 \oplus sl(2,\bbr)_4 \oplus sl(2,\bbr)_6 \oplus sl(2,\bbr)_8 \cr
sl(2,\bbr)_1 \oplus sl(2,\bbr)_4 \oplus sl(2,\bbr)_7 \oplus sl(2,\bbr)_8 \cr
sl(2,\bbr)_1 \oplus sl(2,\bbr)_5 \oplus sl(2,\bbr)_7 \oplus sl(2,\bbr)_8 \cr
sl(2,\bbr)_2 \oplus sl(2,\bbr)_4 \oplus sl(2,\bbr)_6 \oplus sl(2,\bbr)_8 \cr
sl(2,\bbr)_2 \oplus sl(2,\bbr)_4 \oplus sl(2,\bbr)_7 \oplus sl(2,\bbr)_8 \cr
sl(2,\bbr)_2 \oplus sl(2,\bbr)_5 \oplus sl(2,\bbr)_7 \oplus sl(2,\bbr)_8 \cr
so(4,4) \ ,~&~ $\Th = \{2,3,4,8\}$ \cr
so(6,6) \ ,~&~ $\Th = \{2,3,4,5,6,8\}$ \cr
E'_7 \ ,~&~ $\Th = \{1,2,3,4,5,6,8\}$ \cr
}}
All these ~$\cm_\Th\,$, except the last three cases ($so(4,4),so(6,6), E'_7$), have highest/lowest weight representations.

The dimensions of the other factors are, respectively:
\eqn\cuspeightab{
\dim\,\ca_\Th ~=~
\cases{ 8 \cr 7 \cr 6 \cr 5 \cr 4 \cr 4\cr 4\cr 4 \cr 4 \cr 4\cr 4 \cr 4 \cr 2 \cr 1}\ , \qquad
\dim\,\cn^\pm_\Th ~=~
\cases{ 120 \cr 119 \cr  118 \cr 117 \cr 116 \cr 116 \cr 116 \cr 116 \cr 116 \cr 116 \cr 116
 \cr 108 \cr  90 \cr 57}
}

Taking into account \thitb{}
the maximal parabolic subalgebras are determined by:
\eqn\maxeighta{\eqalign{
& \cm^{\rm max}_1   \cong so(7,7)
\ , \qquad \dim\,(\cn^\pm_\Th)^{\rm max} ~=~ 78
\cr
& \cm^{\rm max}_2 \cong sl(7,\bbr) \oplus sl(2,\bbr)
\ , \qquad \dim\,(\cn^\pm_\Th)^{\rm max} ~=~ 98
\cr
& \cm^{\rm max}_3 \cong sl(5,\bbr) \oplus sl(3,\bbr) \oplus sl(2,\bbr)
\ , \qquad \dim\,(\cn^\pm_\Th)^{\rm max} ~=~ 106
\cr
&\cm^{\rm max}_4 \cong sl(5,\bbr) \oplus sl(4,\bbr)
\ , \qquad \dim\,(\cn^\pm_\Th)^{\rm max} ~=~ 104
\cr
& \cm^{\rm max}_5   \cong so(5,5) \oplus sl(3,\bbr)
\ , \qquad \dim\,(\cn^\pm_\Th)^{\rm max} ~=~ 97
\cr
& \cm^{\rm max}_6 \cong E'_6 \oplus sl(2,\bbr)
\ , \qquad \dim\,(\cn^\pm_\Th)^{\rm max} ~=~ 83
\cr
&\cm^{\rm max}_7 \cong  E'_7
\ , \qquad \dim\,(\cn^\pm_\Th)^{\rm max} ~=~ 57
\cr
&\cm^{\rm max}_8 \cong sl(8,\bbr)
\ , \qquad \dim\,(\cn^\pm_\Th)^{\rm max} ~=~ 92
\cr
}}
Clearly, the only  maximal cuspidal parabolic subalgebra is
the one containing ~$\cm^{\rm max}_7\,$.
\nl  The latter case  is also    the only Heisenberg parabolic subalgebra. %25.9.2021

\newsubsec{EIX :~   $E''_8$}

\nt
Another real form of ~$E_8$~ is denoted as ~$E''_8\,$, sometimes as ~$E_{8(-24)}\,$.
The maximal compact subgroup is ~$\ck \cong e_7\oplus su(2)$, ~$\dim_\bbr\,\cp = 112$,
~$\dim_\bbr\,\cn^\pm = 51$.
This real form has discrete series representations.

The split rank is equal to 4, while ~$\cm \cong so(8)$.

The Satake diagram
\eqn\sateightb{
\downcirc{{\a_1}} \riga\black{\a_2} \riga
\black{{\a_3}}\kern-8pt\raise11pt\hbox{$\vert$}
\kern-3.5pt\raise22pt\hbox{$\bullet
{{\scriptstyle{\a_8}}}$} \riga\black{{\a_4}}
\riga\downcirc{{\a_5}} \riga\downcirc{{\a_6}}
\riga\downcirc{{\a_7}}}
Thus, the reduced root system is presented
by a Dynkin-Satake diagram looking like the ~$F_4$~ Dynkin diagram,
cf. \satsixb, but the short roots have multiplicity $8$
(the long - multiplicity $1$).
Going to the $F_4$ diagram we drop the black nodes,
(they give rise to $\cm$), while $\a_1\,,\a_5\,,\a_6\,,\a_7\,$
are mapped to $\l_4\,,\l_3\,,\l_2\,,\l_1\,$, resp., of \satsixb.

Using the above enumeration of $F_4$ simple roots
we shall give the $\cm_\Th$-factors of all
parabolic subalgebras:
\eqn\parabeightb{
\cm_\Th ~=~ \cases{
\cm = so(8)  \ ,~&~ $\Th = \emptyset\ , ~~{\rm minimal}$ \cr
sl(2,\bbr)_j \oplus \cm\ ,~&~ $\Th = \{j\}, ~~~j=1,2$\cr
so(9,1)\ ,~&~ $\Th = \{j\}, ~~~j=3,4$\cr
sl(3,\bbr) \oplus \cm\ ,~&~ $\Th = \{1,2\}$\cr
sl(2,\bbr)_1 \oplus so(9,1)  \ ,~&~ $\Th = \{1,j\}, ~~~j=3,4$\cr
so(10,2) \ ,~&~ $\Th = \{2,3\}$\cr
sl(2,\bbr)_2 \oplus   so(9,1) \ ,~&~ $\Th = \{2,4\}$\cr
e^{iv}_6 \ ,~&~ $\Th = \{3,4\}$\cr
so(11,3)  \ ,~&~ $\Th = \{1,2,3\}$, \cr
sl(3,\bbr) \oplus so(9,1)  \ ,~&~ $\Th = \{1,2,4\}$\cr
sl(2,\bbr)_1\oplus e^{iv}_6  \ ,~&~ $\Th = \{1,3,4\}$\cr
e'''_7 \ ,~&~ $\Th = \{2,3,4\}$\cr
}}
The dimensions of the other factors are, respectively:
\eqn\cuspeightbb{
\dim\,\ca_\Th ~=~
\cases{ 4 \cr 3 \cr 3 \cr 2 \cr 2 \cr 2\cr 2\cr 2 \cr 1 \cr 1\cr 1  \cr 1}\ , \qquad
\dim\,\cn^\pm_\Th ~=~
\cases{ 108 \cr 107 \cr  100 \cr 105 \cr 99 \cr 90 \cr 99 \cr 84 \cr 78 \cr 97 \cr 83\cr 57
  }
}

The maximal parabolic subalgebras are the last four in the above lists
corresponding to ~$\Th^{\rm max}_j\,$, $j=4,3,2,1$, cf. \thitb{}.

The cuspidal ones arise from:
\eqn\cuspeightb{
\cm_\Th ~=~ \cases{
so(8)  \ ,~&~ $\Th = \emptyset\ , ~~{\rm minimal}$ \cr
sl(2,\bbr)_j \oplus so(8)\ ,~&~ $\Th = \{j\}, ~~~j=1,2$\cr
so(10,2) \ ,~&~ $\Th = \{2,3\}$\cr
e'''_7 \ ,~&~ $\Th = \{2,3,4\}$\cr
}}
the last one being also maximal.\nl  The last case  is also    the only Heisenberg parabolic subalgebra. %25.9.2021
\nl
All these ~$\cm_\Th$~ have highest/lowest weight representations.

\newsubsec{FI :~   $F'_4$}

\nt
The split real form of ~$F_4$~ is denoted as ~$F'_4\,$, sometimes as ~$F_{4(+4)}\,$.
The maximal compact subgroup ~$\ck \cong sp(3) \oplus su(2)$, ~$\dim_\bbr\,\cp = 28$,
~$\dim_\bbr\,\cn^\pm = 24$.
This real form has discrete series representations.

Taking into account the enumeration of simple roots as in \satsixb{}
the parabolic subalgebras have $\cm_\Th$-factors as follows:
\eqn\parabfoura{
\cm_\Th ~=~
\cases{
0 ~&~ $\Th = \emptyset$, ~minimal \cr
sl(2,\bbr)_j \ ,~&~ $\Th = \{j\}, ~~~j=1,2,3,4$\cr
sl(3,\bbr)_{jk} \ ,~&~ $\Th = \{j,k\} = \{1,2\}, \{3,4\}$\cr
sl(2,\bbr)_j \oplus sl(2,\bbr)_k\ ,
~&~ $\Th = \{j,k\}\ = \{1,3\}, \{1,4\}, \{2,4\} $ \cr
sp(2,\bbr) \ ,~&~ $\Th  = \{2,3\}$ \cr
so(4,3) \ ,~&~ $\Th = \{1,2,3\}$, \cr
sl(3,\bbr) \oplus sl(2,\bbr) \ ,~&~ $\Th = \{1,2,4\},\{1,3,4\}$, \cr
sp(3,\bbr) \ ,~&~ $\Th  = \{2,3,4\}$ \cr
}}
The dimensions of the other factors are, respectively:
\eqn\parabfourab{
\dim\,\ca_\Th ~=~
\cases{ 4 \cr 3   \cr 2 \cr 2 \cr 2  \cr 1  \cr 1  \cr 1}\ , \qquad
\dim\,\cn^\pm_\Th ~=~
\cases{ 24 \cr 23 \cr  21 \cr 22 \cr 20 \cr 15 \cr 20 \cr 15
  }}

The maximal parabolic subalgebras are in the last three lines in the above lists
corresponding to ~$\Th^{\rm max}_j\,$, $j=4,3,2,1$, cf. \thitb{}.

The cuspidal parabolic subalgebras arise from:
\eqn\cuspsevb{
\cm_\Th ~=~ \cases{
0 ~&~ $\Th = \emptyset$, ~minimal \cr
sl(2,\bbr)_j\  ,~&~ $\Th = \{j\}, ~~~j=1,2,3,4$\cr
sl(2,\bbr)_j \oplus sl(2,\bbr)_k\ ,
~&~ $\Th = \{j,k\}\ = \{1,3\}, \{1,4\}, \{2,4\} $ \cr
sp(2,\bbr) \ ,~&~ $\Th  = \{2,3\}$ \cr
so(4,3) \ ,~&~ $\Th  = \{1,2,3\}$ \cr
sp(3,\bbr) \ ,~&~ $\Th  = \{2,3,4\}$ \cr
}}
the last two being also maximal.
All these ~$\cm_\Th$, except the last but one, have highest/lowest weight representations.\nl  The last case
$sp(3,\bbr)$ is     the only Heisenberg parabolic subalgebra. %25.9.2021

\newsubsec{FII :~   $F''_4$}

\nt
Another real form of ~$F_4$~ is denoted as ~$F''_4\,$, sometimes as ~$F_{4(-20)}\,$.
The maximal compact subgroup ~$\ck \cong so(9)$, ~$\dim_\bbr\,\cp = 16$,
~$\dim_\bbr\,\cn^\pm = 15$.
This real form has discrete series representations.

The split rank is equal to 1, while ~$\cm \cong so(7)$.

The Satake diagram is:
\eqn\satsour{
\black{{\a_1}} \riga\black{{\a_2}}
\Longrightarrow\black{{\a_3}} \riga\downcirc{{\a_4}}}
Thus, the reduced root system is presented
by a Dynkin-Satake diagram looking like the ~$A_1$~ Dynkin diagram
but the roots   have multiplicity~$8$.
Going to the $A_1$ Dynkin diagram we drop the black nodes,
(they give rise to $\cm$), while $\a_4$ becomes
the $A_1$  diagram.

\newsubsec{G :~   $G'_2$}

\nt
The split real form of ~$G_2$~ is denoted as ~$G'_2\,$, sometimes as ~$G_{2(+2)}\,$.
The maximal compact subgroup ~$\ck \cong su(2) \oplus su(2)$, ~$\dim_\bbr\,\cp = 8$,
~$\dim_\bbr\,\cn^\pm = 6$.
This real form has discrete series representations.

The non-minimal parabolic subalgebras have $\cm_\Th$-factors as follows:
\eqn\cuspge{
\cm_\Th ~=~ sl(2,\bbr)_j \ ,~~ \Th = \{j\}, ~~~j=1,2 }
They are cuspidal and maximal.
All  ~$\cm_\Th$~ have highest/lowest weight representations.
\nl  They are also Heisenberg parabolic subalgebras. %25.9.2021

\vskip 10mm

\newsec{Summary and Outlook}

\nt
In the present paper we have started the systematic explicit construction
of the invariant differential operators by giving explicit description of
one of the main ingredients in our setting - the cuspidal parabolic subalgebras.
We explicated also the maximal parabolic subalgebras, since these are important
even when they are not cuspidal. In sequels of this paper \Dobp{} we shall
present the construction of the invariant differential operators
and expand the scheme to the supersymmetric case, in view of applications
to conformal field theory and string theory.

\vskip 10mm

\nt {\bf Acknowledgements.}

\nt The author would like to thank for
hospitality the Abdus Salam International Center for Theoretical
Physics, where part of the work was done. This work was supported in
part by the Bulgarian National Council for Scientific Research,
grant F-1205/02,  the Alexander von Humboldt Foundation in the
framework of the Clausthal-Leipzig-Sofia Cooperation,
 and the European RTN  'Forces-Universe',
contract MRTN-CT-2004-005104. The author would like to thank the
anonymous referee for numerous remarks which contributed to
improving the exposition.

\np

\parskip=4pt
\baselineskip=12pt
\parindent 10pt
\voffset .5cm

\def\tablerule{\noalign{\hrule}}

\centerline{\bf Appendix}
\centerline{\bf Table of Cuspidal Parabolic Subalgebras}
\medskip
\vbox{\offinterlineskip
\halign{\baselineskip12pt
\strut\vrule#\hskip0.1truecm &
#\hfil&
\vrule#\hskip0.1truecm &
#\hfil&
\vrule#\hskip0.1truecm  &
#\hfil&
\vrule#\hskip0.1truecm  &
#\hfil&
\hskip0.1truecm  \vrule#\cr
\tablerule
&&&&&&&&\cr
%%% HEADER ROW
&~ $\cg$
&&~ $\cm_\Th$
&&~  $\dim_\bbr\ca_\Th$
&&~$\dim_\bbr\cn^\pm_\Th$&\cr
&&&&& &&&\cr
\tablerule
&&&&&&&&\cr
%%%  1st ROW
&~ $\cg_\bac$ && ~$u(1) \oplus \cdots \oplus u(1)$ && ~$\ell$ &&  ~$d-\ell$&\cr
&~$\dim_\bac\, \cg_\bac = d$ && ~$\ell$\ \  factors&&&&&\cr
&~$\rank_\bac\, \cg_\bac = \ell$&&&&&&&\cr
&&&&&&&&\cr
\tablerule
&&&&&&&&\cr
&~ $sl(n,\bbr)$
&&~ $\cm_{\Th({\bar n})} =\mathop{\oplus}\limits_{1\leq t \leq k} \ sl(2,\bbr)_{j_t}$
  && ~$n-1-k$    && $~\ha n (n-1) -k$ &\cr
  &&&&&&&&\cr
&&&~ $0\leq k\leq [n/2]$ &&  &&&\cr
&&&~ $j_t < j_{t+1}-1$
&&&&&\cr
&&& ~ $1\leq j_1\,,   ~~~j_k\leq n-1$
&&&&&\cr
&&&~ minimal: ~$k=0$, ~$\cm_\Th=0$ &&~$n-1$    && $~\ha n (n-1) $ &\cr
&&&&&&&&\cr
\tablerule
&&&&&&&&\cr
&~ $su^*(2n)$ && $~ su(2) \oplus \cdots \oplus  su(2)$ &&
~$n-1$&& ~$2n(n-1)$&\cr
&&&~ $n$ factors &&&&&\cr
&&&&&&&&\cr
\tablerule
&           &&&&&&&\cr
&~$su(p,r)$ &&&&&&&\cr
&~$p\geq r$ &&  $su(p-j,r-j)\oplus  u(1)
\oplus \cdots\oplus  u(1)$ &&~$j$ &&~ $j(2(p+r-j)-1)$&\cr
&&&~ $j$ factors, ~$1\leq j< r$ &&&&&\cr
&~$p>r$ &&~ minimal: ~$j=r$~ from above && ~$r$&&~ $r(2p-1)$&\cr
&~$p=r$&&~ minimal: ~u(1) $\oplus \cdots\oplus  u(1)$ &&~$r$ &&
~ $r(2r-1)$ &\cr
&&&~ \phantom{minimal:} ~ $r-1$~ factors && &&&\cr
\tablerule
&&&&&&&&\cr
&~$so(p,r)$&&&&&&&\cr
& ~$p\geq r$&& ~$\cm_{\Th({\bar n}_s)} \oplus so(p-s,r-s)$  && ~$\leq s$&&&\cr
&&&~ $s=1,2,\ldots,r$, ~~~$p+r$~odd &&&&&\cr
&&&~ $s=1,3,\ldots,r$, ~~~$p,r$~odd &&&&&\cr
&&&~ $s=2,4,\ldots,r$, ~~~$p,r$~even &&&&&\cr
&~$p=r$&& ~$\cm_{\Th^o({\bar n})}$  && &&&\cr
& ~$p\geq r$&&~ minimal: ~$so(p-r)$ &&~$r$ && ~$r(p-1)$& \cr
\tablerule}}

\np

\vbox{\offinterlineskip
\halign{\baselineskip12pt
\strut\vrule#\hskip0.1truecm &
#\hfil&
\vrule#\hskip0.1truecm &
#\hfil&
\vrule#\hskip0.1truecm  &
#\hfil&
\vrule#\hskip0.1truecm  &
#\hfil&
\hskip0.1truecm  \vrule#\cr
\tablerule
&&&&&&&&\cr
%%% HEADER ROW
&~ $\cg$
&&~ $\cm_\Th$
&&~  $\dim_\bbr\ca_\Th$
&&~$\dim_\bbr\cn^\pm_\Th$&\cr
&&&&& &&&\cr
\tablerule
&&&&&&&&\cr
%%%  1st ROW
&~$sp(n,\bbr)$ && ~$\cm_{\Th({\bar n}_{s-1})} \oplus sp(n-s,\bbr)$&& ~$\leq s$&&&\cr
&&&~$s=1,\ldots,n$ && &&&\cr
&&&~ minimal: ~$\cm_\Th=0$, ($s=n$) &&~$n$    && $~ n^2  $ &\cr
&&&&&&&&\cr
\tablerule
&&&&&&&&\cr
&~$sp(p,r)$&& $sp(p-j,r-j)\oplus  sp(1)
\oplus \cdots\oplus  sp(1)$ &&~$j$ && ~$j(4p+4r-4j-1)$&\cr
&~$p\geq r$&& ~$j$~ factors, ~$j=1,\ldots,r$&& &&&\cr
&&&~minimal: ~$j=r$ &&&&&\cr
&&&&&&&&\cr
\tablerule
&&&&&&&&\cr
&~$so^*(2n)$ && $so^*(2n -4j) \oplus so(3) \oplus \cdots \oplus so(3)$&& ~$j$&& ~$j(4n-4j-3)$ &\cr
&&& ~$j$ factors, ~$j=1,\ldots,r\equiv [n/2]$ &&&&&\cr
&&& ~minimal: ~$j=r\equiv [n/2]$~  &&&&&\cr
&&&&&&&&\cr
\tablerule
&&&&&&&&\cr
&~EI $\cong E'_6$  && ~0, minimal && ~$6$&& ~$36$&\cr
&&& ~$sl(2,\bbr)_j$ && ~$5$ &&~$35$&\cr
&&& ~$j=1,\ldots,6$ &&&&&\cr
&&& ~$sl(2,\bbr)_j \oplus sl(2,\bbr)_k$&&~$4$&&~$34$&\cr
&&& ~$j+1<k$, ~$\{j,k\} \neq \{3,6\}$;
~$(j,k) =(5,6)$ &&&&&\cr
&&& ~$sl(2,\bbr)_j \oplus sl(2,\bbr)_k \oplus sl(2,\bbr)_\ell$ &&~$3$&&~$33$&\cr
&&& ~$(j,k,\ell) = (1,3,5),(1,4,6),(1,5,6),(2,4,6),(2,5,6)$ &&&&&\cr
&&& ~$so(4,4)$ && ~$2$&&~$24$&\cr
&&&&&&&&\cr
\tablerule
&&&&&&&&\cr
&~EII $\cong E''_6$  && ~$u(1)\oplus u(1)$,  ~minimal && ~$4$&& ~$36$&\cr
&&& ~$sl(2,\bbr)_j \oplus u(1)\oplus u(1)$, ~$j=1,2$&& ~$3$ &&~$35$&\cr
&&& ~$so(4,4)\oplus u(1)$ && ~$1$&&~$24$&\cr
&&&&&&&&\cr
\tablerule
&&&&&&&&\cr
&~EIII $\cong E'''_6$  && ~$so(6)\oplus so(2)$ && ~$2$&& ~$30$&\cr
&&& ~$su(5,1)$ && ~$1$&& ~$21$&\cr
&&&&&&&&\cr
\tablerule
&&&&&&&&\cr
&~EIV $\cong E^{iv}_6$  && ~$so(8)$ && ~$2$&& ~$24$&\cr
\tablerule
}}

\np

\vbox{\offinterlineskip
\halign{\baselineskip12pt
\strut\vrule#\hskip0.1truecm &
#\hfil&
\vrule#\hskip0.1truecm &
#\hfil&
\vrule#\hskip0.1truecm  &
#\hfil&
\vrule#\hskip0.1truecm  &
#\hfil&
\hskip0.1truecm  \vrule#\cr
\tablerule
&&&&&&&&\cr
%%% HEADER ROW
&~ $\cg$
&&~ $\cm_\Th$
&&~  $\dim_\bbr\ca_\Th$
&&~$\dim_\bbr\cn^\pm_\Th$&\cr
&&&&& &&&\cr
\tablerule
&&&&&&&&\cr
%%%  1st ROW
&~EV $\cong E'_7$  && ~0, minimal && ~$7$&& ~$63$&\cr
&&& ~$sl(2,\bbr)_j\,, ~~j=1,\ldots,7$ && ~$6$ &&~$62$&\cr
&&& ~$sl(2,\bbr)_j \oplus sl(2,\bbr)_k$&&~$5$&&~$61$&\cr
&&& ~$j+1<k$, ~$\{j,k\} \neq \{3,7\}$;
~$(j,k) =(6,7)$ &&&&&\cr
&&& ~$sl(2,\bbr)_j \oplus sl(2,\bbr)_k \oplus sl(2,\bbr)_\ell$ &&~$4$&&~$60$&\cr
&&& ~$(j,k,\ell) = (1,3,5),(1,3,6),(1,4,6),(1,4,7),$&&&&&\cr
&&& ~$(1,5,7),(1,6,7),(2,4,6),(2,4,7),(2,5,7),(2,6,7)$ &&&&&\cr
&&& ~$sl(2,\bbr)_1 \oplus sl(2,\bbr)_4 \oplus sl(2,\bbr)_6 \oplus sl(2,\bbr)_7$ && ~$3$&&~$59$&\cr
&&& ~$sl(2,\bbr)_2 \oplus sl(2,\bbr)_4 \oplus sl(2,\bbr)_6 \oplus sl(2,\bbr)_7$ && ~$3$&&~$59$&\cr
&&& ~$so(4,4)$ && ~$3$&&~$51$&\cr
&&& ~$so(6,6)$ && ~$1$&&~$33$&\cr
\tablerule
&&&&&&&&\cr
&~EVI $\cong E''_7$  && ~$su(2)\oplus su(2)\oplus su(2)$,  ~minimal && ~$4$&& ~$60$&\cr
&&& ~$sl(2,\bbr)_j \oplus su(2)\oplus su(2)\oplus su(2)$, ~$j=1,2$&& ~$3$ &&~$59$&\cr
&&& ~$so(6,2)\oplus su(2)$ && ~$2$&&~$50$&\cr
&&& ~$so^*(12)$ && ~$1$&&~$33$&\cr
\tablerule
&&&&&&&&\cr
&~EVII $\cong E'''_7$  && ~$so(8)$,  ~minimal && ~$3$&& ~$51$&\cr
&&& ~$sl(2,\bbr)_3 \oplus so(8)$&& ~$2$ &&~$50$&\cr
&&& ~$so(10,2)$ && ~$1$&&~$33$&\cr
\tablerule
&&&&&&&&\cr
&~FI $\cong F'_4$  && ~$0$,  ~minimal && ~$4$&& ~$24$&\cr
&&& ~$sl(2,\bbr)_j\,,  ~~j=1,2,3,4$&& ~$3$ &&~$23$&\cr
&&& ~$sl(2,\bbr)_j \oplus sl(2,\bbr)_k\,,
~(j,k) = (13),(14),(24)$&&~$2$&&~$22$&\cr
&&& ~$sp(2,\bbr)$&& ~$2$ &&~$20$&\cr
&&& ~$so(4,3)$&& ~$1$ &&~$15$&\cr
&&& ~$sp(3,\bbr)$&& ~$1$ &&~$15$&\cr
\tablerule
&&&&&&&&\cr
&~FII $\cong F''_4$  && ~$so(7)$  && ~$1$&& ~$15$&\cr
\tablerule
&&&&&&&&\cr
&~G $\cong G'_2$  && ~$0$,  ~minimal && ~$2$&& ~~$6$&\cr
&&& ~$sl(2,\bbr)_j\,,  ~~j=1,2$&& ~$1$ &&~~$5$&\cr
&&&&&&&&\cr
\tablerule
}}

\np

\vbox{\offinterlineskip
\halign{\baselineskip12pt
\strut\vrule#\hskip0.1truecm &
#\hfil&
\vrule#\hskip0.1truecm &
#\hfil&
\vrule#\hskip0.1truecm  &
#\hfil&
\vrule#\hskip0.1truecm  &
#\hfil&
\hskip0.1truecm  \vrule#\cr
\tablerule
&&&&&&&&\cr
%%% HEADER ROW
&~ $\cg$
&&~ $\cm_\Th$
&&~  $\dim_\bbr\,\ca_\Th$
&&~$\dim_\bbr\,\cn^\pm_\Th$&\cr
&&&&& &&&\cr
\tablerule
&&&&&&&&\cr
%%%  1st ROW
&~EVIII $\cong E'_8$  && ~0, minimal && ~$8$&& ~$120$&\cr
&&& ~$sl(2,\bbr)_j\,, ~~j=1,\ldots,8$ && ~$7$ &&~$119$&\cr
&&& ~$sl(2,\bbr)_j \oplus sl(2,\bbr)_k$&&~$6$&&~$118$&\cr
&&& ~$j+1<k$, ~$\{j,k\} \neq \{3,8\}$;
~$(j,k) =(7,8)$ &&&&&\cr
&&& ~$sl(2,\bbr)_j \oplus sl(2,\bbr)_k \oplus sl(2,\bbr)_\ell$ &&~$5$&&~$117$&\cr
&&& ~$(j,k,\ell) = (1,3,5),(1,3,6),(1,3,7),(1,4,6),$&&&&&\cr
&&& ~$(1,4,7),(1,4,8),(1,5,7),(1,5,8),(1,6,8),$&&&&&\cr
&&& ~$(2,4,6),(2,4,7),(2,4,8),(2,5,7),(2,5,8),$&&&&&\cr
&&& ~$(2,6,8),(2,7,8),(3,5,7),(4,6,8),(4,7,8),(5,7,8),$ &&&&&\cr
&&& ~$sl(2,\bbr)_j \oplus sl(2,\bbr)_k \oplus sl(2,\bbr)_\ell \oplus sl(2,\bbr)_m$ && ~$4$&&~$116$&\cr
&&& ~$(j,k,\ell,m) = (1,3,5,7),(1,4,6,8),(1,4,7,8),$&&&&&\cr
&&& ~$(1,5,7,8),(2,4,6,8),(2,4,7,8),(2,5,7,8)$&&&&&\cr
&&& ~$so(4,4)$ && ~$4$&&~$108$&\cr
&&& ~$so(6,6)$ && ~$2$&&~~$90$&\cr
&&& ~$E'_7$ && ~$1$&&~~$57$&\cr
&&&&&&&&\cr
\tablerule
&&&&&&&&\cr
&~EIX $\cong E''_8$  && ~$so(8)$,  ~minimal && ~$4$&& ~$108$&\cr
&&& ~$sl(2,\bbr)_j \oplus so(8), ~~j=1,2$&& ~$3$ &&~$107$&\cr
&&& ~$so(10,2)$ && ~$2$&&~~$90$&\cr
&&& ~$E'_7$ && ~$1$&&~~$57$&\cr
&&&&&&&&\cr
\tablerule }}

\np

\parskip=0pt
\listrefs

\end